\documentclass{article}

\usepackage{arxiv}

\usepackage[utf8]{inputenc} 
\usepackage[T1]{fontenc}    
\usepackage{hyperref}       
\usepackage{url}            
\usepackage{booktabs}       
\usepackage{amsfonts}       
\usepackage{nicefrac}       
\usepackage{microtype}      
\usepackage{lipsum}
\usepackage{times}
\usepackage{epsfig}
\usepackage{graphicx}
\usepackage{amsmath}
\usepackage{amssymb}
\usepackage{caption,setspace}
\usepackage{subfig}
\usepackage{cite}
\usepackage{CJK}
\usepackage{float}
\usepackage{color}
\usepackage{verbatim}
\usepackage{indentfirst}
\title{Adaptive Backstepping Control for Fractional-Order Nonlinear Systems with External Disturbance and Uncertain Parameters Using Smooth Control}

\author{
  Xinyao~Li \\
    School of Electrical and Electronic Engineering\\
Nanyang Technological University\\
Singapore, 639798\\
\texttt{E180209@e.ntu.edu.sg}\\
   \And
 Changyun~Wen \\
  School of Electrical and Electronic Engineering\\
Nanyang Technological University\\
Singapore, 639798\\
  \texttt{ecywen@ntu.edu.sg} \\
  \And
  Ying~Zou \\
  School of Electrical and Electronic Engineering\\
Nanyang Technological University\\
Singapore, 639798\\
\texttt{zouy0011@e.ntu.edu.sg} \\
}

\begin{document}
\maketitle

\begin{abstract}
In this paper, we consider controlling a class of single-input-single-output (SISO) commensurate fractional-order nonlinear systems with parametric uncertainty and external disturbance. Based on backstepping approach, an adaptive controller is proposed with adaptive laws that are used to estimate the unknown system parameters and the bound of unknown disturbance. Instead of using discontinuous functions such as the $\mathrm{sign}$ function, an auxiliary function is employed to obtain a smooth control input that is still able to achieve perfect tracking in the presence of bounded disturbances. Indeed, global boundedness of all closed-loop signals and asymptotic perfect tracking of fractional-order system output to a given reference trajectory are proved by using fractional directed Lyapunov method. To verify the effectiveness of the proposed control method, simulation examples are presented.
\end{abstract}


\section{Introduction}
Fractional calculus owns a history of more than 300 years. It is a branch of mathematics that deals with non-integer order derivatives and integrals. Compared with integer order calculus, fractional order integral and derivative can both be treated as weighted integral and thus they have the properties of hereditary and infinite memory \cite{kilbas2006theory,podlubny1994fractional}, which can also be seen from their definitions given in Section 2. Such properties give the most significant meaning of fractional-order derivative compared to integer-order derivative that does not have such properties. Thus, using fractional order models could better and more accurately describe the characteristics of the real world systems than integer order models, as elaborated in \cite{torvik1984appearance,caputo1971new,friedrich1991relaxation}. In the past few years, many researchers have paid significant attention to fractional order calculus and constructed models for real-world systems, including viscoelasticity, complex systems, neural networks, transmission line, multi agent systems, and so on, see for examples \cite{nigmatullin2006recognition,koeller1984applications,yu2012alpha,wang1987realizations,liu2018exponential}. Moreover, fractional order systems as well as their controls have also been studied in recent years \cite{hartley2002dynamics,chen2006robust,chen2009fractional}. However, it is difficult to simply apply the approaches of controller design and analysis developed for integer-order systems to fractional-order systems due to the lack of appropriate mathematical tools. For example, the fractional-order derivative of a composite function is the sum of infinite number of terms, which is different from the concise closed form expression of its integer-order derivative easily obtained by applying the chain rule.

Backstepping technique, which demonstrates a step-by-step design procedure by constructing Lyapunov function and virtual control signal at each step, is widely applied in controlling integer-order nonlinear systems, see \cite{wang2017adaptive} and \cite{baleanu2011fractional} for example. However, it is not easy to directly employ backstepping control technique on fractional-order systems due to the challenge of obtaining the fractional derivative of the quadratic-type Lyapunov function. In \cite{baleanu2011fractional}, the backstepping technique is extended to fractional order systems without entirely considering uncertainties. Integer-order Lyapunov method is applied in \cite{baleanu2011fractional} by proving $z\dot{z}<0$ if $zz^{(\alpha)}<0$, where $z$ denotes virtual error defined in backstepping algorithm and $0<\alpha<1$. A method that transforms backstepping control problem for the fractional-order systems to integer-order by taking into account the frequency distributed model is proposed in \cite{wei2015novel, sheng2017adaptive, wei2016adaptive}. In this way, the stability analysis for fractional-order systems is also carried out with integer-order Lyapunov method.

In order to handle disturbances in fractional-order nonlinear systems, nonlinear disturbances observers are designed to counteract the effects caused by unknown disturbances in \cite{pashaei2016new} and \cite{chen2016disturbance}. However, strong assumptions that the disturbances must be constant or they should have bounded derivatives should be met to achieve satisfactory performances. 
Adaptive controllers utilizing $\mathrm{sign}$ function are developed in \cite{liu2017adaptive, song2018adaptive, wang2013synchronization, gong2018adaptive}, which ensure the stabilization/tracking error converges to zero with non-continuous control because of the non-continuity of $\mathrm{sign}$ function. For the purpose of avoiding calculating the fractional-order derivatives of virtual control signals, the involved derivatives are directly subtracted in the design of virtual controllers in subsequent steps in \cite{wang2016adaptive,ding2015non,ding2014adaptive}. However, the boundedness of signals in the closed-loop control systems is not theoretically shown in these works.
Fuzzy logic is employed in \cite{liu2017adaptive} and \cite{ma2019adaptive} to deal with system uncertainties as well as computing fractional order
derivatives of virtual control signals with the assumption that the errors resulted from fuzzy  approximation of the true values of system uncertainties and the fractional order derivatives of virtual control signals are bounded. In addition, both works can only show that the output tracking errors tend to an arbitrary small region. 
Besides, \cite{ma2019adaptive} combines dynamic surface control with adaptive backstepping method for SISO fractional-order nonlinear systems with uncertain system functions and multiple external unknown disturbances. Although it can eliminate the chattering phenomenon in control signals by getting rid of using the $\mathrm{sign}$ function, their results can only be obtained if the initial conditions are located in certain range. Dynamic surface control integrated with neuro-fuzzy network for fractional-order nonlinear system subjects to input constraint is addressed in \cite{song2019neuro}. Similarly, the output tracking error is only guaranteed to converge to a small region around the origin with constraint of the initial conditions. Therefore, it is still open and challenging to construct an adaptive backstepping controller with smooth control signal for fractional nonlinear systems involving both parametric uncertainties and external time-varying disturbance, which ensures asymptotic stability and perfect tracking property without restrictions on the initial conditions.

Inspired by the discussions above, in this paper we address such an issue. To have a smooth control signal, an auxiliary function is used in lieu of the $\mathrm{sign}$ function in the designed controller. We employ the fractional directed Lyapunov method for our design and analysis. As we know, when backstepping approach is employed to design controller for high-order nonlinear systems, the time derivatives of virtual control signals are required and they can easily be obtained for integer-order systems by chain rule. But, as mentioned above, this is not the case for fractional-order systems. To overcome this difficulty, a novel approach for approximating the fractional-order time derivatives of virtual control signals is proposed and adaptive laws are designed to estimate the bounds of approximation errors. It is shown that, with the designed controller and adaptive laws, the closed loop system is globally stable and its output tracks a given reference input asymptotically, even in the presence of external time-varying disturbance and uncertain parameters. To the best of the authors’ knowledge, this is the first paper to have such results. Simulation studies illustrate the effectiveness of the proposed control scheme and also reveal its advantages compared to an existing approach. In summary, the main contribution of this paper is to design a smooth control that achieves asymptotic perfect output tracking/stabilization for fractional-order nonlinear commensurate systems and ensures global stability in the sense that all signals in the closed-loop systems remain globally bounded, even in the presence of unknown bounded external disturbances and also system uncertainties. 

The rest of the paper is organized as follows. Preliminaries and the fractional-order system description are provided in Section \ref{sec:Preliminaries}. In Section \ref{sec:controller}, the design of an adaptive controller is presented in detail. In Section \ref{sec:simulation}, the scheme is illustrated by simulation studies with comparison to that in \cite{liu2017adaptive}. Finally the paper is concluded in Section \ref{sec:conclusion}.

\section{Preliminaries and Problem Formulation}
\label{sec:Preliminaries}


\subsection{Preliminaries}
\emph{Definition 1 \cite{podlubny1998fractional}:} The fractional integral of an integrable function $f(t)$ with $\alpha\in\mathbb{R}^{+}$ is
\begin{equation}
{}_{t_0}\!{\mathit{I}}_t^{\alpha}f(t) = \frac{1}{\Gamma(\alpha)}\int_{t_0}^{t}\frac{f(\tau)}{(t-\tau)^{1-\alpha}}\,d\tau
\label{eq:eqI}
\end{equation}
where ${}_{t_0}\!{\mathit{I}}_t^{\alpha}$ means the fractional integral of order $\alpha$ with initial time ${t_0}$ and $\Gamma(\cdot)$ denotes the well-known Gamma function, which is defined as $\Gamma(z)=\int_{0}^{\infty}e^{-t}t^{z-1}\,dt$, where $z \in \mathbb{C}$. One of the significant properties of Gamma function is \cite{xue2017fractional}: $\Gamma(z+1) = z\Gamma(z),\, \Gamma(n+1) = n\Gamma(n)=n(n-1)\Gamma(n-1)=\cdots=n!,\, \Gamma(-n)= \infty$, where $n \in \mathbb{N}_0=\{n|n \ge 0,n \in \mathbb{N}\}$.

\emph{Definition 2 \cite{podlubny1998fractional}:} The Caputo fractional derivative of a function is shown as
\begin{equation}
{}_{t_0}^{C}\!{\mathcal{D}}_t^{\alpha}f(t) = {}_{t_0}{\mathit{I}}_t^{(m-\alpha)}\frac{\mathrm{d}^m}{\mathrm{d}t^m}f(t)=\frac{1}{\Gamma(m-\alpha)}\int_{t_0}^{t} \frac{f^{(m)}(\tau)}{(t-\tau)^{\alpha-m+1}}\,d\tau
\label{eq:eqCd}
\end{equation}
where $m - 1 < \alpha < m \in\mathbb{Z}^+$. From equation (\ref{eq:eqCd}) we can observe that the Caputo derivative of a constant is $0$. Another commonly used fractional derivative is named Riemann-Liouville (RL) and the RL fractional derivative of a function $f(t)$ is denoted as ${}_{t_0}^{RL}\!{\mathcal{D}}_t^{\alpha}f(t)$. Different from the Caputo derivative, RL derivative of a constant is not equal to $0$ \cite{Lakshmikantham2009TheoryOF,podlubny1998fractional}.

To obtain the unique solution for fractional differential equation ${}_{t_0}\!{\mathcal{D}}_t^{\alpha}x(t)=f(x,t)$, ($m - 1 < \alpha < m \in\mathbb{Z}^+$ and $t\ge t_0$), the initial values need to be determined. According to \cite{li2007remarks,podlubny1998fractional} and \cite{bandyopadhyay2015stabilization}, fractional differential equations with Caputo-type derivative have initial values that are in-line with integer-order differential equations, i.e.\ $x(t_0),x'(t_0),\dots,x^{(m-1)}(t_0)$, which contain specific physical interpretations. On the contrary, although mathematically the initial value problem for RL fractional differential equations is rigorous and solvable, it lacks of practical explanation since the physical meanings of these initial conditions are unknown yet. Therefore, Caputo-type fractional systems are frequently employed in practical analysis. 

\emph{Lemma 1 \cite{aguila2014lyapunov}:} Assume $x(t)\in \mathbb{R}^n$ is a smooth function, $n \in \mathbb{Z}^+$, then for $\forall t \ge t_0$,
\begin{equation}
\frac{1}{2}{}_{t_0}^{C}\!{\mathcal{D}}_t^{\alpha}[x^{\mathrm{T}}(t)\mathit{P}x(t)] \leq x^{\mathrm{T}}(t)\mathit{P}{}_{t_0}^{C}\!{\mathcal{D}}_t^{\alpha}x(t),\quad 0 < \alpha < 1
\label{eq:eqL1}
\end{equation}
where $\mathit{P} \in \mathbb{R}^{n\times n}$ is a positive definite constant matrix.

\emph{Lemma 2 \cite{podlubny1998fractional}:} If $0 < \alpha < 2$, $\beta$ represents an arbitrary complex number and real number $\mu$ satisfies that $\frac{\pi\alpha}{2} < \mu < \mathrm{min}\{\pi,\pi\alpha\}$, then for any integer $n \ge 1$, 
\begin{equation}
\mathrm{E}_{\alpha,\beta}(z) = -\sum_{j=1}^n\frac{z^{-j}}{\Gamma(\beta-\alpha j)}+\mathrm{o}(\left|z\right|^{-1-n}),
\label{eq:eqL3}
\end{equation}
where $\mu \leq \left|\mathrm{arg}(z)\right| \leq \pi$ and $\left|z\right| \rightarrow \infty$. Here $\mathrm{E}_{\alpha,\beta}(z)$ is the Mittag-Leffler function, which is described as follows:
\begin{equation}
\mathrm{E}_{\alpha,\beta}(z) = \sum_{k=0}^\infty\frac{z^k}{\Gamma(\alpha k+\beta)}, \ (\alpha > 0, \, \beta > 0).
\label{eq:eqMittag}
\end{equation}

\emph{Lemma 3 \cite{li2007remarks,aguila2014lyapunov}:} If the Caputo fractional derivative ${}_{t_0}^{C}\!{\mathcal{D}}_t^{\alpha}f(t)$ is integrable, then
\begin{equation}
{}_{t_0}\!{\mathit{I}}_t^{\alpha}{}_{t_0}^{C}\!{\mathcal{D}}_t^{\alpha}f(t) = f(t) - \sum_{k=0}^{m-1}\frac{f^{(k)}(t_0)}{k!}(t-{t_0})^k
\label{eq:eqL4_1}
\end{equation}
where $m - 1 < \alpha < m \in\mathbb{Z}^+$. Particularly, for $0 < \alpha \leq 1$, we can obtain
\begin{equation}
{}_{t_0}\!{\mathit{I}}_t^{\alpha}{}_{t_0}^{C}\!{\mathcal{D}}_t^{\alpha}f(t) = f(t) - f(t_0).
\label{eq:eqL4_2}
\end{equation}

\emph{Definition 3 \cite{li2009mittag,li2010stability, zhang2011asymptotical}:} For fractional nonautonomous system ${}_{t_0}\!{\mathcal{D}}_t^{\alpha}x_i(t)=f_i(x,t)$, where $i = 1,2,\dots,n$, $0 < \alpha < 1$, initial condition is $x(t_0) = [x_1(t_0),x_2(t_0),\dots,x_n(t_0)]^\mathrm{T} \in \mathbb{R}^n$, ${}_{t_0}\!{\mathcal{D}}_t^{\alpha}$ indicates Caputo or RL fractional derivative, $f_i(x,t): [t_0, \infty) \times \Omega \rightarrow \mathbb{R}^n$ is locally Lipschitz in $x$ and piecewise continuous in $t$ (which insinuates the existence and uniqueness of the solution to the fractional systems \cite{podlubny1998fractional}) and $\Omega \in \mathbb{R}^n$ stands for a region that contains the origin $x = [0,0,\dots,0]^\mathrm{T}$. The equilibrium $x^* =[x_1^*,x_2^*,\dots,x_n^*]^\mathrm{T}$ of this system is defined as ${}_{t_0}\!{\mathcal{D}}_t^{\alpha}x^*=f_i(x^*,t)$ for $t \ge t_0$. Without loss of generality, we set the equilibrium $x = [0,0,\dots,0]^\mathrm{T}$.

\emph{Lemma 4 \cite{gallegos2015fractional}:} Assume $f(t): \mathbb{R}^{+} \rightarrow \mathbb{R}^{+}$ is uniformly continuous and $\lim_{t \rightarrow \infty}{}_{t_0}\!{\mathit{I}}_t^{\alpha}f(t) =0$ with $0 < \alpha < 1$ and $\forall t > t_0 > 0$, then $\lim_{t \rightarrow \infty}f(t) = 0$.

\emph{Lemma 5 \cite{zhang2017new}:} If $x(t): [t_0, \infty) \times \mathbb{R}^n \rightarrow \mathbb{R}^n$ is uniformly continuous, ${}_{t_0}\!{\mathit{I}}_t^{\alpha}\left|x(t)\right|^p \leq \mathit{K}$ for $\forall t > t_0 > 0$ and $0 < \alpha < 1$ with constants $p,\,\mathit{K} > 0$, then $\lim_{t\rightarrow \infty}x(t)=0$.

\emph{Lemma 6 \cite{zuo2014adaptive}:} The following inequality holds for $x \in \mathbb{R}$:
\begin{equation}
\left|x \right| < \epsilon + xsg(x,\epsilon)
\label{eq:eqL2_1}
\end{equation}
where 
\begin{equation}
sg(x,\epsilon)=\frac{x}{\sqrt{x^2+{\epsilon}^2}}
\label{eq:eqL2_2}
\end{equation}
and $\epsilon$ is a differentiable function which meets $\epsilon > 0$ and $\int_{0}^{t}\epsilon(\tau)d\tau < \infty, \; \forall t \ge 0$.

From equation (\ref{eq:eqL2_2}), it can be derived that $sg(x,\epsilon)$ is differentiable and thus it is a smooth function for $\forall t \ge 0$.

\subsection{Problem Formulation}
In this paper, Caputo-type definition of the fractional derivatives is utilized. Consider the following class of $n$-th order SISO uncertain commensurate fractional-order nonlinear systems with time-varying disturbance in strict feedback form:
\begin{equation}
\begin{cases}
\begin{aligned}
{}_0^{C}\!{\mathcal{D}}_t^{\alpha}&x_1(t) = x_2(t)+\phi_1^\mathrm{T}(\overline{x}_1(t))\theta_1\\
{}_0^{C}\!{\mathcal{D}}_t^{\alpha}&x_2(t) = x_3(t)+\phi_2^\mathrm{T}(\overline{x}_2(t))\theta_2 \\
&\vdots\\
{}_0^{C}\!{\mathcal{D}}_t^{\alpha}&x_{n-1}(t) = x_n(t)+\phi_{n-1}^\mathrm{T}(\overline{x}_{n-1}(t))\theta_{n-1} \\
{}_0^{C}\!{\mathcal{D}}_t^{\alpha}&x_{n}(t) = g(x(t))u(t)+\phi_{n}^\mathrm{T}(x(t))\theta_{n}+d(t)\\
y(t)&=x_1(t)\\
\end{aligned}
\end{cases}
\label{eq:eqPro}
\end{equation}
where the fractional-orders of all the state equations are equal to $\alpha \in (0, 1]$, $u(t) \in \mathbb{R}$ and $y(t) \in \mathbb{R}$ represent the control input and system output respectively, $x(t)=[x_1(t),x_2(t),\cdots,x_n(t)]^\mathrm{T} \in \mathbb{R}^n$ denotes the measurable state vector, $\overline{x}_i(t)=[x_1(t),x_2(t),\cdots,x_i(t)]^\mathrm{T} \in\mathbb{R}^i$, $\phi_i(\overline{x}_{i}(t)) \in \mathbb{R}^p$ ($i=1,2,\cdots,n$) is a vector with its elements being known nonlinear smooth functions, $\theta_{i} \in \mathbb{R}^p$ ($i=1,2,\cdots,n$) is an unknown constant vector, $g(x(t))$ is a known non-zero smooth nonlinear function and $d(t)$ stands for an unknown bounded time-varying external disturbance with unknown bound $D$.

\emph{Remark 1:} Different from the control strategies in \cite{yang2019h_,zong2019composite,sun2018disturbance,zhang2017fault} that are designed with disturbance observer or disturbance resilience, a fractional order adaptive law, which can be found in (\ref{eq:eqALn}), is to be derived to estimate the unknown upper bound $D$ of the disturbance. The external time-varying disturbance $d(t)$ is allowed to be norm-bounded without any condition on its derivative in this paper. By employing the estimated bound in designing control input $u$, such unknown disturbance can be compensated, which helps us to solve the following control problem.

The control problem is to design an adaptive controller for the class of systems described in (\ref{eq:eqPro}) such that the following objectives are achieved: 1) the closed-loop system is globally stable in the sense that all the signals including parameter estimates are bounded; 2) the system output $y(t)$ asymptotically tracks a reference signal $r(t)$, i.e.\ $\lim_{t\rightarrow \infty}[y(t)-r(t)]=0$.

\emph{Assumption 1:} The given reference signal $r(t)$ and its $\alpha$-th order Caputo-type fractional derivative ${}_0^{C}\!{\mathcal{D}}_t^{\alpha}r(t)$ are smooth and bounded.

\section{Adaptive Controller Design and Stability Analysis}
\label{sec:controller}
\subsection{Adaptive Controller Design}
To achieve the above objectives, an adaptive controller is designed based on the backstepping design procedure. Defining virtual errors as follows: $z_1=x_1 - r$, which is also known as tracking error, $z_i = x_i - \alpha_{i-1} \,(i=2,\dots,n)$, where $\alpha_j\,(j=1,2,\dots,n-1)$ denotes the virtual control signal that will be designed iteratively later.

\emph{Step 1:} Consider the first Lyapunov function $\overline{\mathit{V}}_1 = \frac{1}{2}{z_1}^2$. According to Lemma 1, the Caputo derivative of $\overline{\mathit{V}}_1$ is
\begin{equation}
\begin{aligned}
{}_0^{C}\!{\mathcal{D}}_t^{\alpha}\overline{\mathit{V}}_1 &= \frac{1}{2}{}_0^{C}\!{\mathcal{D}}_t^{\alpha}{z_1}^2 \leq {z_1}{}_0^{C}\!{\mathcal{D}}_t^{\alpha}{z_1} = z_1[{}_0^{C}\!{\mathcal{D}}_t^{\alpha}{x_1}-{}_0^{C}\!{\mathcal{D}}_t^{\alpha}{r}] = z_1[x_2+\phi_1^\mathrm{T}(x_1)\theta_1-{}_0^{C}\!{\mathcal{D}}_t^{\alpha}{r}] \\
&= z_1[z_2+\alpha_1+\phi_1^\mathrm{T}(x_1)\theta_1-{}_0^{C}\!{\mathcal{D}}_t^{\alpha}{r}]= z_1z_2+z_1\alpha_1+z_1\phi_1^\mathrm{T}(x_1)\theta_1-z_1{}_0^{C}\!{\mathcal{D}}_t^{\alpha}{r}.\\
\end{aligned}
\label{eq:eqVbar1_1}
\end{equation}

Let virtual control signal $\alpha_1$ be
\begin{equation}
\alpha_1 = -c_1z_1-\phi_1^\mathrm{T}(x_1)\hat{\theta}_1+{}_0^{C}\!{\mathcal{D}}_t^{\alpha}{r},
\label{eq:eqa1}
\end{equation}
where $\hat{\theta}_1$ stands for the estimates of $\theta_1$ and $c_1$ is a positive design parameter. Defining $\tilde{\theta}_1 = \theta_1 - \hat{\theta}_1$, then we have ${}_0^{C}\!{\mathcal{D}}_t^{\alpha}\tilde{\theta}_1 = - {}_0^{C}\!{\mathcal{D}}_t^{\alpha}\hat{\theta}_1$. Therefore, inequality (\ref{eq:eqVbar1_1}) becomes
\begin{equation}
\begin{aligned}
{}_0^{C}\!{\mathcal{D}}_t^{\alpha}\overline{\mathit{V}}_1 \leq &z_1z_2+z_1\alpha_1+z_1\phi_1^\mathrm{T}(x_1)\theta_1-z_1{}_0^{C}\!{\mathcal{D}}_t^{\alpha}{r}=z_1z_2-c_1{z_1}^2-z_1\phi_1^\mathrm{T}(x_1)\hat{\theta}_1+z_1{}_0^{C}\!{\mathcal{D}}_t^{\alpha}{r}+z_1\phi_1^\mathrm{T}(x_1)\theta_1-z_1{}_0^{C}\!{\mathcal{D}}_t^{\alpha}{r}\\
=&z_1z_2-c_1{z_1}^2+z_1\phi_1^\mathrm{T}(x_1)\tilde{\theta}_1.\\
\end{aligned}
\label{eq:eqVbar1_2}
\end{equation}

Let $\mathit{V}_1=\overline{\mathit{V}}_1+\frac{1}{2}\tilde{\theta}_1^\mathrm{T}\Gamma_1^{-1}\tilde{\theta}_1$, where $\Gamma_1$ is positive definite matrix, then the Caputo derivative of $\mathit{V}_1$ is
\begin{equation}
\begin{aligned}
{}_0^{C}\!{\mathcal{D}}_t^{\alpha}\mathit{V}_1 =&{}_0^{C}\!{\mathcal{D}}_t^{\alpha}\overline{\mathit{V}}_1+\frac{1}{2}{}_0^{C}\!{\mathcal{D}}_t^{\alpha}\tilde{\theta}_1^\mathrm{T}\Gamma_1^{-1}\tilde{\theta}_1
\leq{}_0^{C}\!{\mathcal{D}}_t^{\alpha}\overline{\mathit{V}}_1+\tilde{\theta}_1^\mathrm{T}\Gamma_1^{-1}{}_0^{C}\!{\mathcal{D}}_t^{\alpha}\tilde{\theta}_1
\leq z_1z_2-c_1{z_1}^2+z_1\phi_1^{\mathrm{T}}(x_1)\tilde{\theta}_1+\tilde{\theta}_1^\mathrm{T}\Gamma_1^{-1}{}_0^{C}\!{\mathcal{D}}_t^{\alpha}\tilde{\theta}_1\\
=&z_1z_2-c_1{z_1}^2+z_1\phi_1^{\mathrm{T}}(x_1)\tilde{\theta}_1-\tilde{\theta}_1^\mathrm{T}\Gamma_1^{-1}{}_0^{C}\!{\mathcal{D}}_t^{\alpha}\hat{\theta}_1.\\
\end{aligned}
\label{eq:eqV1_1}
\end{equation}
Designing adaptive law as
\begin{equation}
\begin{aligned}
{}_0^{C}\!{\mathcal{D}}_t^{\alpha}\hat{\theta}_1 &= z_1\Gamma_1\phi_1(x_1),\\
\end{aligned}
\label{eq:eqAL_1}
\end{equation}
then inequality (\ref{eq:eqV1_1}) becomes
\begin{equation}
\begin{aligned}
{}_0^{C}\!{\mathcal{D}}_t^{\alpha}\mathit{V}_1 \leq&z_1z_2-c_1{z_1}^2+z_1\phi_1^{\mathrm{T}}(x_1)\tilde{\theta}_1-\tilde{\theta}_1^{\mathrm{T}}\Gamma_1^{-1}\Gamma_1\phi_1(x_1)z_1
=z_1z_2-c_1{z_1}^2.
\end{aligned}
\label{eq:eqV1_2}
\end{equation}

\emph{Step $i$ ($i=2,\cdots,n-1$):} The fractional order derivative of $z_i$ is
\begin{equation}
\begin{aligned}
{}_0^{C}\!{\mathcal{D}}_t^{\alpha}z_i=&{}_0^{C}\!{\mathcal{D}}_t^{\alpha}x_i-{}_0^{C}\!{\mathcal{D}}_t^{\alpha}\alpha_{i-1}=x_{i+1}+\phi_i^\mathrm{T}(\overline{x}_i(t))\theta_i-{}_0^{C}\!{\mathcal{D}}_t^{\alpha}\alpha_{i-1}=z_{i+1}+\alpha_i+\phi_i^\mathrm{T}(\overline{x}_i(t))\theta_i-{}_0^{C}\!{\mathcal{D}}_t^{\alpha}\alpha_{i-1}.
\end{aligned}
\label{eq:eqzder1}
\end{equation}

Inspired by the idea in \cite{jumarie2006modified} and \cite{wang2017fractional} where ${}_0^{}\!{\mathcal{D}}_x^{\alpha}f[u(x)]$ is approximated by $f_u'(u){}_0^{}\!{\mathcal{D}}_x^{\alpha}u(x)$ with approximation error $\rho(t)$ which is bounded by an unknown bound $\bar{\rho}$, ${}_0^{C}\!{\mathcal{D}}_t^{\alpha}\alpha_{i-1}$ is approximated by $\sum_{j=1}^{i-1}(\frac{\partial\alpha_{i-1}}{\partial {x_{j}}}){}_0^{C}\!{\mathcal{D}}_t^{\alpha}x_j+\sum_{j=1}^{i-1}(\frac{\partial\alpha_{i-1}}{\partial {\hat{\theta}_{v,j}}}){}_0^{C}\!{\mathcal{D}}_t^{\alpha}\hat{\theta}_{v,j}$. Therefore by defining $\rho_{i-1}(t)$ as the approximation error with bound $\bar{\rho}_{i-1}$, (\ref{eq:eqzder1}) can be written as
\begin{equation}
{}_0^{C}\!{\mathcal{D}}_t^{\alpha}z_i=z_{i+1}+\alpha_i+\phi_{v,i}^{\mathrm{T}}\theta_{v,i}+\zeta_i+\rho_{i-1}.
\label{eq:eqzder2}
\end{equation}

The functions and parameters in (\ref{eq:eqzder2}) are given below
\begin{equation}
\begin{aligned}
\phi_{v,i}=&[\phi_i^{\mathrm{T}}(\overline{x}_i),(-\frac{\partial\alpha_{i-1}}{\partial {x_{i-1}}})\phi_{i-1}^{\mathrm{T}}(\overline{x}_{i-1}),\cdots,(-\frac{\partial\alpha_{i-1}}{\partial {x_2}})\phi_2^{\mathrm{T}}(\overline{x}_2),(-\frac{\partial\alpha_{i-1}}{\partial {x_1}})\phi_1^{\mathrm{T}}(\overline{x}_1)]^{\mathrm{T}},\\
\theta_{v,i}=&[\theta_{i}^{\mathrm{T}},\theta_{i-1}^{\mathrm{T}},\cdots,\theta_1^{\mathrm{T}}]^{\mathrm{T}},\\
\zeta_i=&\sum_{j=2}^{i}(-\frac{\partial\alpha_{i-1}}{\partial {x_{j-1}}})x_j+\sum_{j=1}^{i-1}(-\frac{\partial\alpha_{i-1}}{\partial {\hat{\theta}_{v,j}}}){}_0^{C}\!{\mathcal{D}}_t^{\alpha}\hat{\theta}_{v,j},\\
\end{aligned}
\label{eq:eqzder_vars}
\end{equation}
where $\hat{\theta}_{v,j}$ is the estimate of $\theta_{v,j},\,j=1,2,\cdots,i-1$.

Choose virtual control signal $\alpha_i$ as
\begin{equation}
\alpha_i=-z_{i-1}-c_iz_i-\phi_{v,i}^{\mathrm{T}}\hat{\theta}_{v,i}-\zeta_i-\hat{\rho}_{i-1},
\label{eq:eqai}
\end{equation}
where $c_i$ denotes a positive design parameter, $\hat{\theta}_{v,i}$ means the estimate of $\theta_{v,i}$ and $\hat{\rho}_{i-1}$ is the estimation of $\bar{\rho}_{i-1}$ with their adaptive law respectively designed as
\begin{equation}
\begin{aligned}
{}_0^{C}\!{\mathcal{D}}_t^{\alpha}\hat{\theta}_{v,i}&=z_i\Gamma_i\phi_{v,i},\\
{}_0^{C}\!{\mathcal{D}}_t^{\alpha}\hat{\rho}_{i-1}&=z_i\lambda_{i-1},
\end{aligned}
\label{eq:eqALi}
\end{equation}
where $\Gamma_i$ is positive definite and $\lambda_{i-1} > 0$. Let $\mathit{V}_i=\mathit{V}_{i-1}+\frac{1}{2}{z_i}^2+\frac{1}{2}\tilde{\theta}_{v,i}^{\mathrm{T}}\Gamma_i^{-1}\tilde{\theta}_{v,i}+\frac{1}{2\lambda_{i-1}}{\tilde{\rho}_{i-1}}^2$, where $\tilde{\theta}_{v,i}=\theta_{v,i}-\hat{\theta}_{v,i}$ and $\tilde{\rho}_{i-1}=\bar{\rho}_{i-1}-\hat{\rho}_{i-1}$, then 
\begin{equation}
\begin{aligned}
{}_0^{C}\!{\mathcal{D}}_t^{\alpha}\mathit{V}_i \leq&{}_0^{C}\!{\mathcal{D}}_t^{\alpha}\mathit{V}_{i-1}+z_i{}_0^{C}\!{\mathcal{D}}_t^{\alpha}z_i+\tilde{\theta}_{v,i}^{\mathrm{T}}\Gamma_i^{-1}{}_0^{C}\!{\mathcal{D}}_t^{\alpha}\tilde{\theta}_{v,i}+\frac{1}{\lambda_{i-1}}\tilde{\rho}_{i-1}{}_0^{C}\!{\mathcal{D}}_t^{\alpha}\tilde{\rho}_{i-1}\\
=&{}_0^{C}\!{\mathcal{D}}_t^{\alpha}\mathit{V}_{i-1}+z_i{}_0^{C}\!{\mathcal{D}}_t^{\alpha}z_i-\tilde{\theta}_{v,i}^{\mathrm{T}}\Gamma_i^{-1}{}_0^{C}\!{\mathcal{D}}_t^{\alpha}\hat{\theta}_{v,i}-\frac{1}{\lambda_{i-1}}\tilde{\rho}_{i-1}{}_0^{C}\!{\mathcal{D}}_t^{\alpha}\hat{\rho}_{i-1}.
\end{aligned}
\label{eq:eqVi_1}
\end{equation}

Since ${}_0^{C}\!{\mathcal{D}}_t^{\alpha}\mathit{V}_{i-1} \leq z_{i-1}z_i-\sum_{j=1}^{i-1}c_j{z_j}^2$, thus we get
\begin{equation}
{}_0^{C}\!{\mathcal{D}}_t^{\alpha}\mathit{V}_i \leq z_iz_{i+1}-\sum_{j=1}^{i}c_j{z_j}^2.
\label{eq:eqVi_2}
\end{equation}

\emph{Step $n$:} The fractional order derivative of $z_n$ is
\begin{equation}
{}_0^{C}\!{\mathcal{D}}_t^{\alpha}z_n=g(x)u+d+\phi_{v,n}^{\mathrm{T}}\theta_{v,n}+\zeta_n+\rho_{n-1},
\label{eq:eqznder}
\end{equation}
where
\begin{equation}
\begin{aligned}
\phi_{v,n}=&[\phi_n^{\mathrm{T}}(x),(-\frac{\partial\alpha_{n-1}}{\partial {x_{n-1}}})\phi_{n-1}^{\mathrm{T}}(\overline{x}_{n-1}),\cdots,(-\frac{\partial\alpha_{n-1}}{\partial {x_2}})\phi_2^{\mathrm{T}}(\overline{x}_2),(-\frac{\partial\alpha_{n-1}}{\partial {x_1}})\phi_1^{\mathrm{T}}(\overline{x}_1)]^{\mathrm{T}},\\
{\theta}_{v,n}=&[{\theta}_{n}^{\mathrm{T}},{\theta}_{n-1}^{\mathrm{T}},\cdots,{\theta}_1^{\mathrm{T}}]^{\mathrm{T}},\\
\zeta_n=&\sum_{j=2}^n(-\frac{\partial\alpha_{n-1}}{\partial {x_{j-1}}})x_j+\sum_{j=1}^{n-1}(-\frac{\partial\alpha_{n-1}}{\partial {\hat{\theta}_{v,j}}}){}_0^{C}\!{\mathcal{D}}_t^{\alpha}\hat{\theta}_{v,j},\\
\rho_{n-1}=& \sum_{j=1}^{n-1}(\frac{\partial\alpha_{n-1}}{\partial {x_{j}}}){}_0^{C}\!{\mathcal{D}}_t^{\alpha}x_j+\sum_{j=1}^{n-1}(\frac{\partial\alpha_{n-1}}{\partial {\hat{\theta}_{v,j}}}){}_0^{C}\!{\mathcal{D}}_t^{\alpha}\hat{\theta}_{v,j}\\&-{}_0^{C}\!{\mathcal{D}}_t^{\alpha}\alpha_{n-1},\\
\bar{\rho}_{n-1} \ge& \left|\rho_{n-1}\right|.
\end{aligned}
\label{eq:eqznder_vars}
\end{equation}

Considering the Lyapunov function $\overline{\mathit{V}}_n =\mathit{V}_{n-1}+\frac{1}{2}{z_n}^2$, then the Caputo derivative of $\overline{\mathit{V}}_n$ is
\begin{equation}
\begin{aligned}
{}_0^{C}\!{\mathcal{D}}_t^{\alpha}\overline{\mathit{V}}_n =& {}_0^{C}\!{\mathcal{D}}_t^{\alpha}\mathit{V}_{n-1}+\frac{1}{2}{}_0^{C}\!{\mathcal{D}}_t^{\alpha}{z_n}^2 
\leq {}_0^{C}\!{\mathcal{D}}_t^{\alpha}\mathit{V}_{n-1}+{z_n}{}_0^{C}\!{\mathcal{D}}_t^{\alpha}{z_n} \\
\leq& z_{n-1}z_{n}-\sum_{j=1}^{n-1}c_j{z_j}^2+z_n[g(x)u+d+\phi_{v,n}^{\mathrm{T}}\theta_{v,n}+\zeta_n +\bar{\rho}_{n-1}] \\
\leq& z_{n-1}z_{n}-\sum_{j=1}^{n-1}c_j{z_j}^2+z_n[g(x)u+\phi_{v,n}^{\mathrm{T}}\theta_{v,n}+\zeta_n+\bar{\rho}_{n-1}]+\left|z_n\right|D,\\
\end{aligned}
\label{eq:eqVbarn_1}
\end{equation}
where $D$ is an unknown bound of $d(t)$. Considering Lemma 6, since $\left|z_n \right| < \epsilon + z_nsg(z_n,\epsilon)$, hence $\left|z_n \right|D \leq \epsilon D + z_nsg(z_n,\epsilon)D$. Therefore, (\ref{eq:eqVbarn_1}) can be written as
\begin{equation}
\begin{aligned}
{}_0^{C}\!{\mathcal{D}}_t^{\alpha}\overline{\mathit{V}}_n \leq& z_{n-1}z_{n}-\sum_{j=1}^{n-1}c_j{z_j}^2+z_n[g(x)u+\phi_{v,n}^{\mathrm{T}}\theta_{v,n}+\zeta_n+\bar{\rho}_{n-1}]+\epsilon D + z_nsg(z_n,\epsilon)D.
\end{aligned}
\label{eq:eqVbarn_2}
\end{equation}

Let $\mathit{V}_n=\mathit{V}_{n-1}+\frac{1}{2}{z_n}^2+\frac{1}{2}\tilde{\theta}_{v,n}^{\mathrm{T}}\Gamma_n^{-1}\tilde{\theta}_{v,n}+\frac{1}{2\lambda_{n-1}}\tilde{\rho}_{n-1}^2+\frac{1}{2\eta}\tilde{D}^2$, where $\Gamma_n$ is positive definite, $\eta > 0$, $\lambda_{n-1} > 0$, $\tilde{\theta}_{v,n}=\theta_{v,n}-\hat{\theta}_{v,n}$, $\tilde{\rho}_{n-1}=\bar{\rho}_{n-1}-\hat{\rho}_{n-1}$, $\tilde{D}=D-\hat{D}$ and $\hat{\theta}_{v,n}$, $\hat{\rho}_{n-1}$ and $\hat{D}$ represent the estimates of $\theta_{v,n}$, $\bar{\rho}_{n-1}$ and $D$ respectively. Then 
\begin{equation}
\begin{aligned}
{}_0^{C}\!{\mathcal{D}}_t^{\alpha}\mathit{V}_n \leq&
{}_0^{C}\!{\mathcal{D}}_t^{\alpha}\mathit{V}_{n-1}+z_n{}_0^{C}\!{\mathcal{D}}_t^{\alpha}z_n+\tilde{\theta}_{v,n}^{\mathrm{T}}\Gamma_n^{-1}{}_0^{C}\!{\mathcal{D}}_t^{\alpha}\tilde{\theta}_{v,n}+\frac{1}{\lambda_{n-1}}\tilde{\rho}_{n-1}{}_0^{C}\!{\mathcal{D}}_t^{\alpha}\tilde{\rho}_{n-1}+\frac{1}{\eta}\tilde{D}{}_0^{C}\!{\mathcal{D}}_t^{\alpha}\tilde{D}\\
=&{}_0^{C}\!{\mathcal{D}}_t^{\alpha}\mathit{V}_{n-1}+z_n{}_0^{C}\!{\mathcal{D}}_t^{\alpha}z_n-\tilde{\theta}_{v,n}^{\mathrm{T}}\Gamma_n^{-1}{}_0^{C}\!{\mathcal{D}}_t^{\alpha}\hat{\theta}_{v,n}-\frac{1}{\lambda_{n-1}}\tilde{\rho}_{n-1}{}_0^{C}\!{\mathcal{D}}_t^{\alpha}\hat{\rho}_{n-1}-\frac{1}{\eta}\tilde{D}{}_0^{C}\!{\mathcal{D}}_t^{\alpha}\hat{D}.
\end{aligned}
\label{eq:eqVn_1_1}
\end{equation}
According to (\ref{eq:eqVbarn_1}) and (\ref{eq:eqVbarn_2}), we then have
\begin{equation}
\begin{aligned}
{}_0^{C}\!{\mathcal{D}}_t^{\alpha}\mathit{V}_n\leq&z_{n-1}z_{n}-\sum_{j=1}^{n-1}c_j{z_j}^2+z_n[g(x)u+\phi_{v,n}^{\mathrm{T}}\theta_{v,n}+\zeta_n+\bar{\rho}_{n-1}]+\epsilon D + z_nsg(z_n,\epsilon)D\\
&-\tilde{\theta}_{v,n}^{\mathrm{T}}\Gamma_n^{-1}{}_0^{C}\!{\mathcal{D}}_t^{\alpha}\hat{\theta}_{v,n}-\frac{1}{\lambda_{n-1}}\tilde{\rho}_{n-1}{}_0^{C}\!{\mathcal{D}}_t^{\alpha}\hat{\rho}_{n-1}-\frac{1}{\eta}\tilde{D}{}_0^{C}\!{\mathcal{D}}_t^{\alpha}\hat{D}.
\end{aligned}
\label{eq:eqVn_1}
\end{equation}

Finally the real control input $u$ is designed as
\begin{equation}
u=\frac{1}{g(x)}[-z_{n-1}-c_nz_n-\phi_{v,n}^{\mathrm{T}}\hat{\theta}_{v,n}-sg(z_n,\epsilon)\hat{D}-\zeta_n-\hat{\rho}_{n-1}],
\label{eq:equn}
\end{equation}
where function $\epsilon(t)=e^{-at}$ with $a$ being a positive constant and $c_n$ is positive design parameter.

Also we have the following adaptive laws designed
\begin{equation}
\begin{aligned}
{}_0^{C}\!{\mathcal{D}}_t^{\alpha}\hat{\theta}_{v,n}&=z_n\Gamma_n\phi_{v,n},\\
{}_0^{C}\!{\mathcal{D}}_t^{\alpha}\hat{\rho}_{n-1}&=z_n\lambda_{n-1},\\
{}_0^{C}\!{\mathcal{D}}_t^{\alpha}\hat{D}&=\eta z_nsg(z_n,\epsilon).
\end{aligned}
\label{eq:eqALn}
\end{equation}
Hence by substituting (\ref{eq:equn}) and (\ref{eq:eqALn}) into (\ref{eq:eqVn_1}), we have
\begin{equation}
{}_0^{C}\!{\mathcal{D}}_t^{\alpha}\mathit{V}_n \leq -\sum_{j=1}^{n}c_j{z_j}^2+\epsilon D.
\label{eq:eqVn_2}
\end{equation}

\emph{Remark 2:} Function $sg(z_n,\epsilon)$ in the control law (\ref{eq:equn}) is engaged to compensate for the effects of the unknown external time-varying disturbance. Due to the fact that $z_nd\leq \lvert z_n\rvert D$ shown in (\ref{eq:eqVbarn_1}), such compensation is achieved by handling its bound $D$. Furthermore, to employ $sg(z_n,\epsilon)$ in the control input for the compensation, we choose function $\epsilon(t)=e^{-at} > 0$ where $a>0$. Since this function satisfies $\int_{0}^{t}\epsilon(\tau)d\tau < \infty, \; \forall t \ge 0$, it, together with other parts of the proposed adaptive controller including the estimated $\hat{D}$ obtained from the adaptive law in (\ref{eq:eqALn}), ensures that perfect asymptotic output tracking is achieved in our work, which is proved in the main result. Also since $sg(z_n,\epsilon)$ is differentiable, the obtained control signal in (\ref{eq:equn}) is smooth. 

\emph{Remark 3:} Different from some existing control schemes such as that in \cite{zhang2017fault}, we do not estimate the actual disturbance for achieving the control objectives pointed out in the problem formulation in Section 2. Instead, we estimate the upper bound $D$ of the disturbance. In addition, the adaptive laws in (\ref{eq:eqALn}), including the law for updating $\hat{D}$, and control law (\ref{eq:equn}) are derived from (\ref{eq:eqVn_1}) in such a way that (\ref{eq:eqVn_2}) is obtained. Our main results in the next subsection are proved based on (\ref{eq:eqVn_2}). As seen from the proof, $\hat{D}$ is not required to converge to the true unknown bound $D$ and, in fact, such a convergence is not one of the control objectives.

\subsection{Main Result}
For the adaptive controller design above, we have the following result.

\emph{Theorem 1:}  Consider the closed-loop system consisting of fractional system (\ref{eq:eqPro}) and adaptive controller with control law (\ref{eq:equn}) and adaptive laws (\ref{eq:eqAL_1}), (\ref{eq:eqALi}) and (\ref{eq:eqALn}). The system is globally stable in the sense that all signals in the closed-loop system are uniformly bounded and also, asymptotic tracking is achieved, i.e.\ $\lim_{t\rightarrow \infty}[y(t)-r(t)]=0$.

\emph{Proof:} Taking fractional integration of both sides of inequality (\ref{eq:eqVn_2}) gives
\begin{equation}
\begin{aligned}
\mathit{V}_n(t) \leq& {}_{t_0}^{C}\!{\mathit{I}}_t^{\alpha}(-\sum_{j=1}^{n}c_j{z_j}^2) + {}_{t_0}^{C}\!{\mathit{I}}_t^{\alpha}(\epsilon D) + \mathit{V}_n(0) 
\leq {}_{t_0}^{C}\!{\mathit{I}}_t^{\alpha}(\epsilon D) +\mathit{V}_n(0). 
\end{aligned}
\label{eq:eqVnless1}
\end{equation}

In (\ref{eq:eqVnless1}), the first term on the right hand side can be computed as follow: 
\begin{equation}
\begin{aligned}
{}_{t_0}^{C}\!{\mathit{I}}_t^{\alpha}(\epsilon D) &=\frac{D}{\Gamma(\alpha)}\int_0^t\frac{\epsilon(\tau)}{(t-\tau)^{1-\alpha}}d\tau =\frac{D}{\Gamma(\alpha)}\int_0^t\frac{e^{-a\tau}}{(t-\tau)^{1-\alpha}}d\tau
=Dt^\alpha\mathrm{E}_{1,\,\alpha+1}(-at).
\end{aligned}
\label{eq:eqT2pro}
\end{equation}

According to Lemma 2, since $\left|\mathrm{arg}(-at)\right|=\pi$ and $\left|-at\right| \rightarrow \infty$ for $ t \rightarrow \infty$, by choosing integer $n=1$, then we have
\begin{equation}
\begin{aligned}
\mathrm{E}_{1,\,\alpha+1}(-at) =& -\frac{(-at)^{-1}}{\Gamma(\alpha+1-1)}+\mathrm{o}(\left|-at\right|^{-2})
=\frac{a^{-1}}{\Gamma(\alpha)t}+\mathrm{o}\big(\frac{1}{\left|-at\right|^{2}}\big).
\end{aligned}
\label{eq:eqEo}
\end{equation}
Hence following (\ref{eq:eqEo}),
\begin{equation}
\begin{aligned}
t^\alpha\mathrm{E}_{1,\,\alpha+1}(-at) = \frac{a^{-1}}{\Gamma(\alpha)t^{1-\alpha}}+t^\alpha\mathrm{o}\big(\frac{1}{\left|-at\right|^{2}}\big).
\end{aligned}
\label{eq:eqtEo}
\end{equation}

As $t \rightarrow \infty$, the  right hand side of (\ref{eq:eqtEo}) tends to $0$. Therefore we have $\lim_{t \rightarrow \infty}t^{\alpha}\mathrm{E}_{1,\,\alpha+1}(-at) = 0$, which implies $\lim_{t \rightarrow \infty}{}_{t_0}^{C}\!{\mathit{I}}_t^{\alpha}(\epsilon D)$ is also $0$. Hence from Lemma 4, we get $\lim_{t \rightarrow \infty}(\epsilon D)=0$. Finally we can come to the conclusion that $\mathit{V}_n(t)$ is bounded from (\ref{eq:eqVnless1}), thus every signal in $\mathit{V}_n(t)$ is bounded. Then all the virtual control signals $\alpha_j\,(j=1,2,\dots,n-1)$ are bounded in accordance with (\ref{eq:eqa1}) and (\ref{eq:eqai}) and further $x_i \,(i=1,2,\dots,n)$ are also bounded. Besides, from (\ref{eq:equn}) it can be noticed that $u$ is bounded. Therefore, from system (\ref{eq:eqPro}) it can be shown that the derivatives of $x_i$ exist and are bounded, which reveals that $x_i$ are uniformly continuous.

Since $r$ is uniformly continuous according to Assumption 1, hence $z_1=x_1 - r$ is also uniformly continuous. As a result, $\alpha_1$ is uniformly continuous from (\ref{eq:eqa1}) and (\ref{eq:eqAL_1}). Subsequently, as $z_i = x_i - \alpha_{i-1} \,(i=2,\dots,n)$, $z_i$ can be proved recursively that are uniformly continuous. From (\ref{eq:eqVnless1}) we know that ${}_{t_0}^{C}\!{\mathit{I}}_t^{\alpha}(\sum_{j=1}^{n}c_j{z_j}^2)$ is bounded. According to Lemma 5, we have $\lim_{t\rightarrow \infty} z_j = 0,\,j=1,2,\cdots,n$. Therefore the output $y(t)$ tracks reference signal $r(t)$ asymptotically.



\emph{Remark 4:} If the control objective is to globally  asymptotically stabilize the system, it could be ensured with the same procedure by treating $r(t)=0$. In addition, if $\alpha = 1$, then the corresponding results in this paper become those for integer-order case.

\section{Simulation Results}
\label{sec:simulation}
In this section a second-order and a third-order fractional nonlinear systems are presented as examples for demonstrating and comparing the proposed control method with an existing scheme in \cite{liu2017adaptive}.
\subsection{A Second-order Example}
The system to be controlled is given as follows
\begin{equation}
\begin{cases}
\begin{aligned}
{}_0^{C}\!{\mathcal{D}}_t^{\alpha}&x_1(t) = x_2(t)+\phi_1(x_1(t))\theta_1\\
{}_0^{C}\!{\mathcal{D}}_t^{\alpha}&x_{2}(t) = u(t)+\phi_{2}(x_1(t),x_2(t))\theta_{2}+d(t)\\
y(t)&=x_1(t)\\
\end{aligned}
\end{cases}
\label{eq:eq2nd}
\end{equation}
where $\alpha =0.95,\;\phi_1(x_1)=-0.4x_1^2,\;\phi_{2}(x_1,x_2)=-0.1x_2+\frac{x_2-0.5x_1^2}{1+x_1^4}$. Here $\theta_1=\theta_2=1$ and external time-varying disturbance $d(t)=\mathrm{sin}(t)+\mathrm{cos}(t)+2\mathrm{U}(t-15)$ where $\mathrm{U}(t)$ is the unit step function are used for simulation purpose, but unknown to designer. The term $2\mathrm{U}(t-15)$ implies that disturbance $d(t)$ has a step jump at $t=15s$ and thus it does not have bounded derivative.

\begin{figure}[!htb]
\begin{tabular}{cc}
\begin{minipage}[t]{0.48\linewidth}
\includegraphics[width=7.4cm,height=5.4cm]{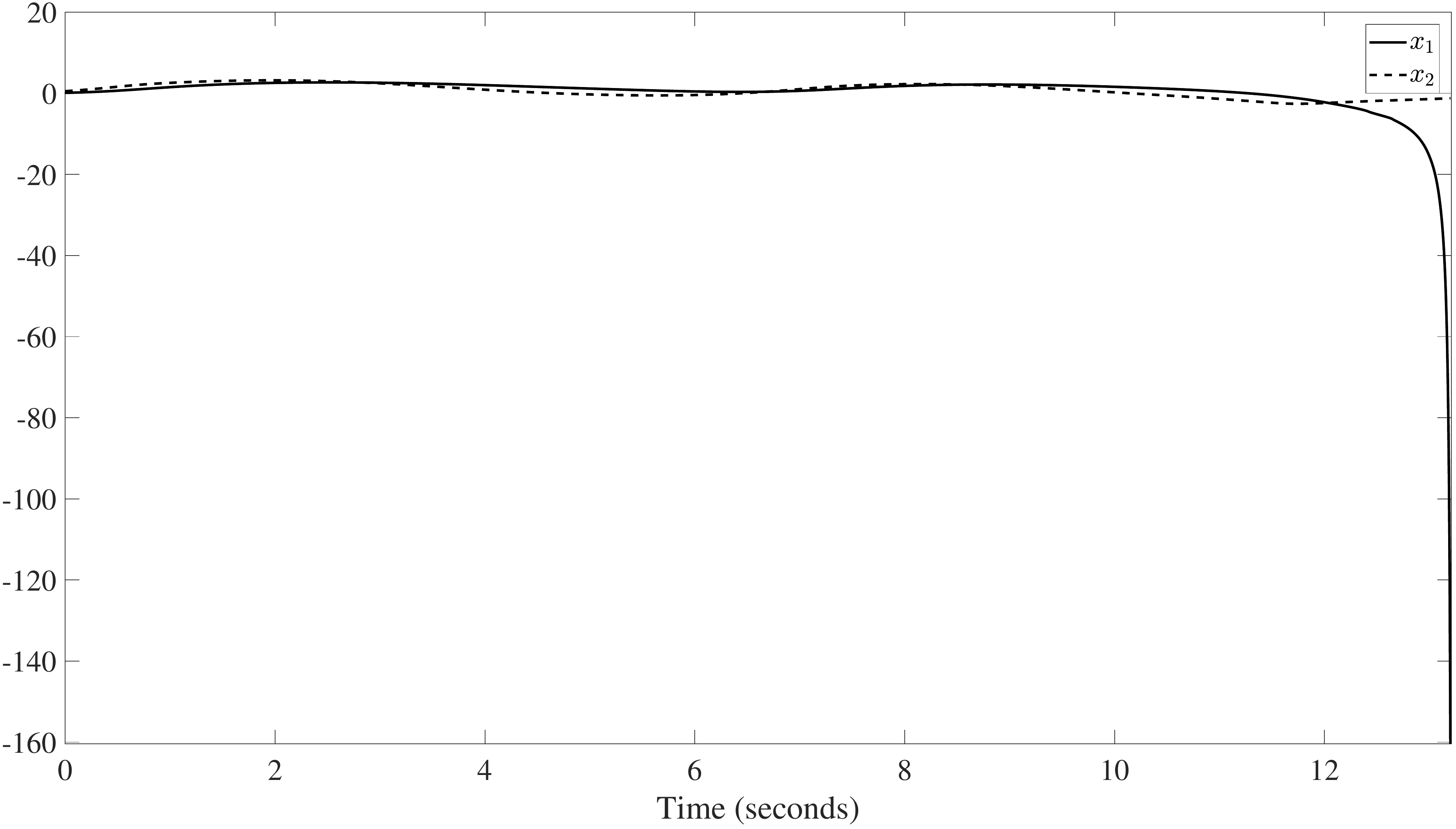}
\caption{The state variables of uncontrolled system (\ref{eq:eq2nd}).}
\label{fig:x1x2nou}
\end{minipage}
\begin{minipage}[t]{0.48\linewidth}
\includegraphics[width=7.4cm,height=5.4cm]{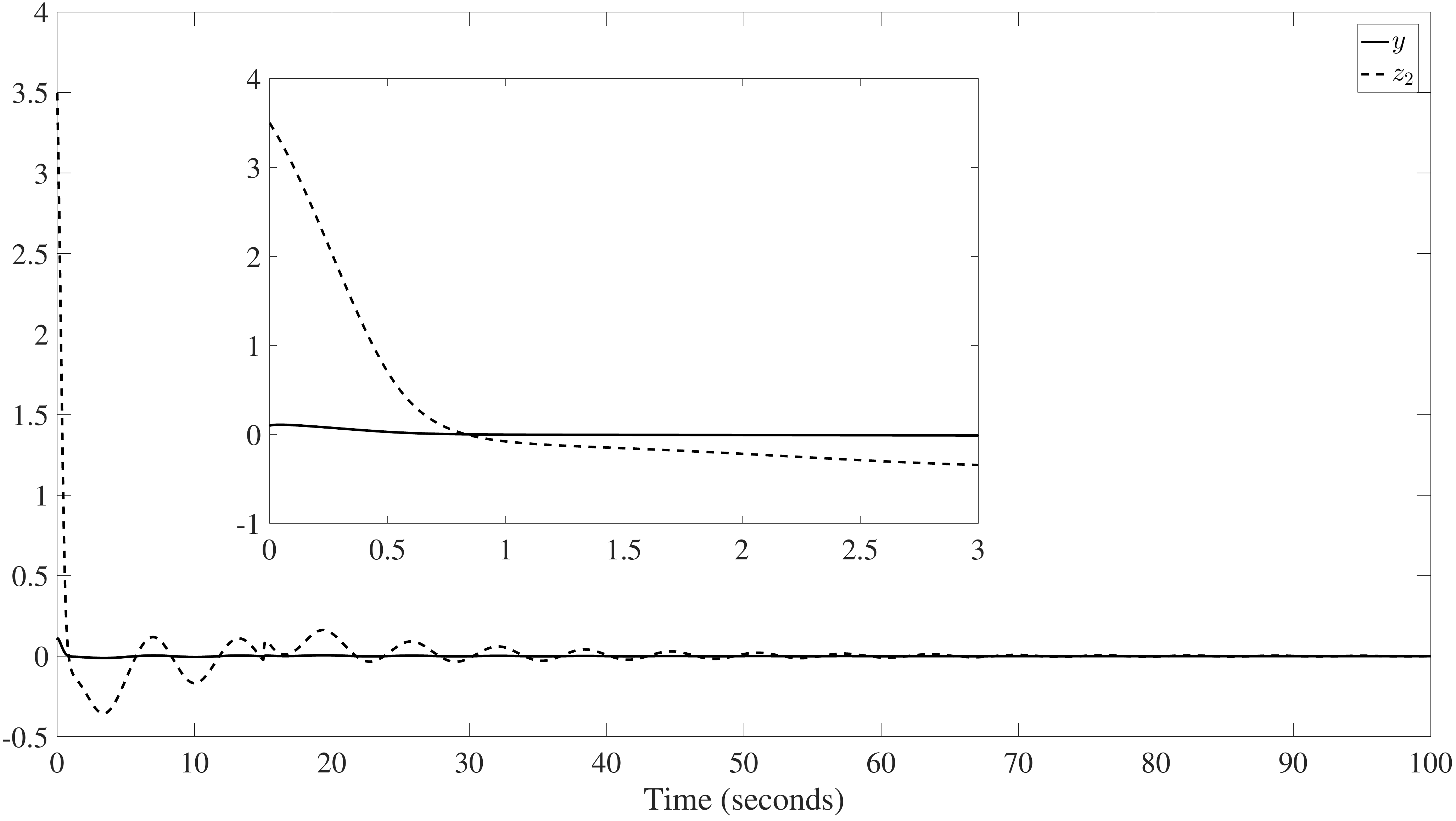}
\caption{The system output $y=x_1=z_1$ and virtual error $z_2$.}
\label{fig:yz2_d}
\end{minipage}
\end{tabular}
\end{figure}

\begin{figure}[h]
\begin{tabular}{cc}
\begin{minipage}[t]{0.48\linewidth}
\includegraphics[width=7.4cm,height=5.4cm]{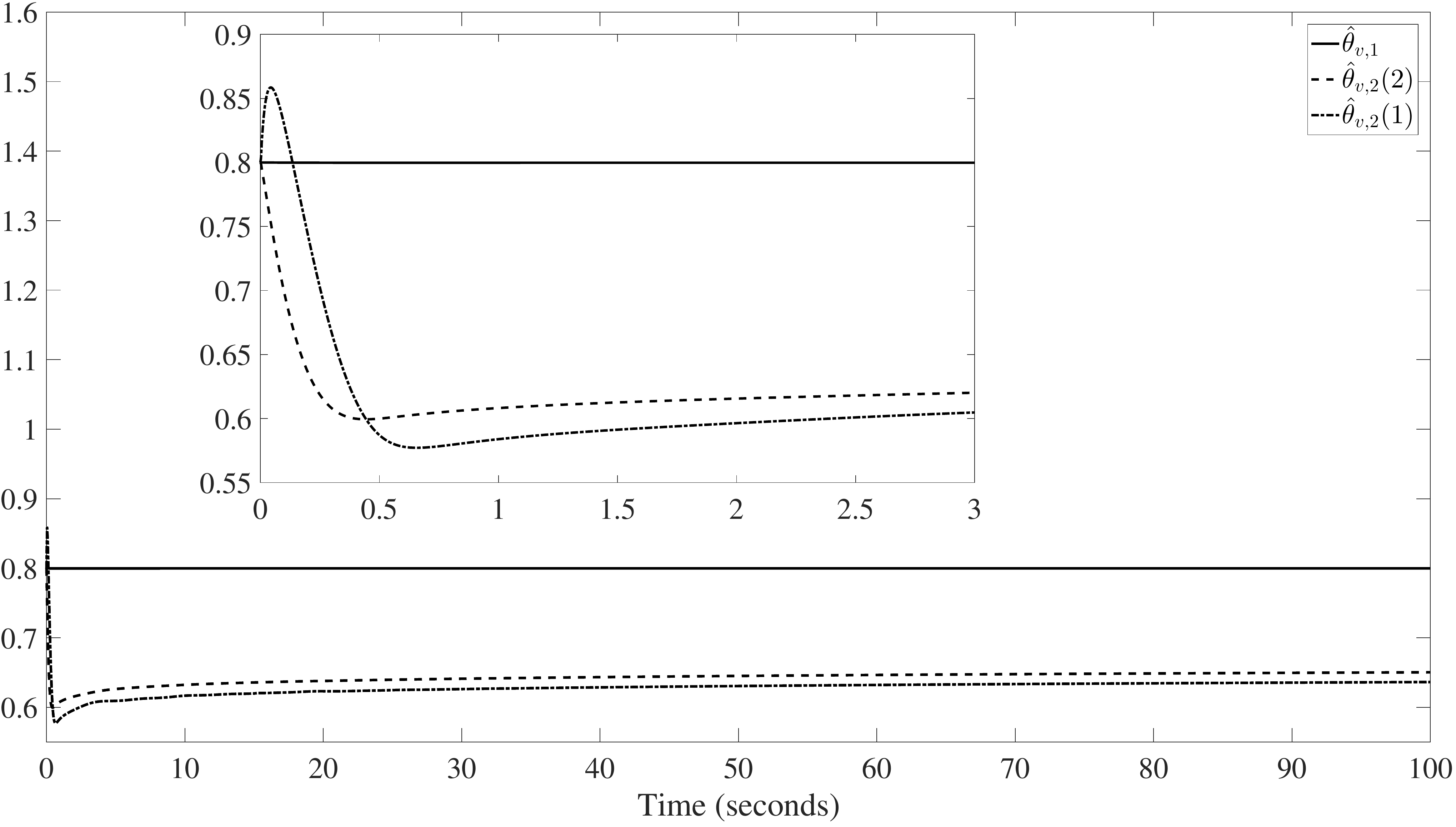}
\caption{The estimates $\hat{\theta}_{v,1}$ and $\hat{\theta}_{v,2}$.}
\label{fig:thetas_d}
\end{minipage}
\begin{minipage}[t]{0.48\linewidth}
\includegraphics[width=7.4cm,height=5.4cm]{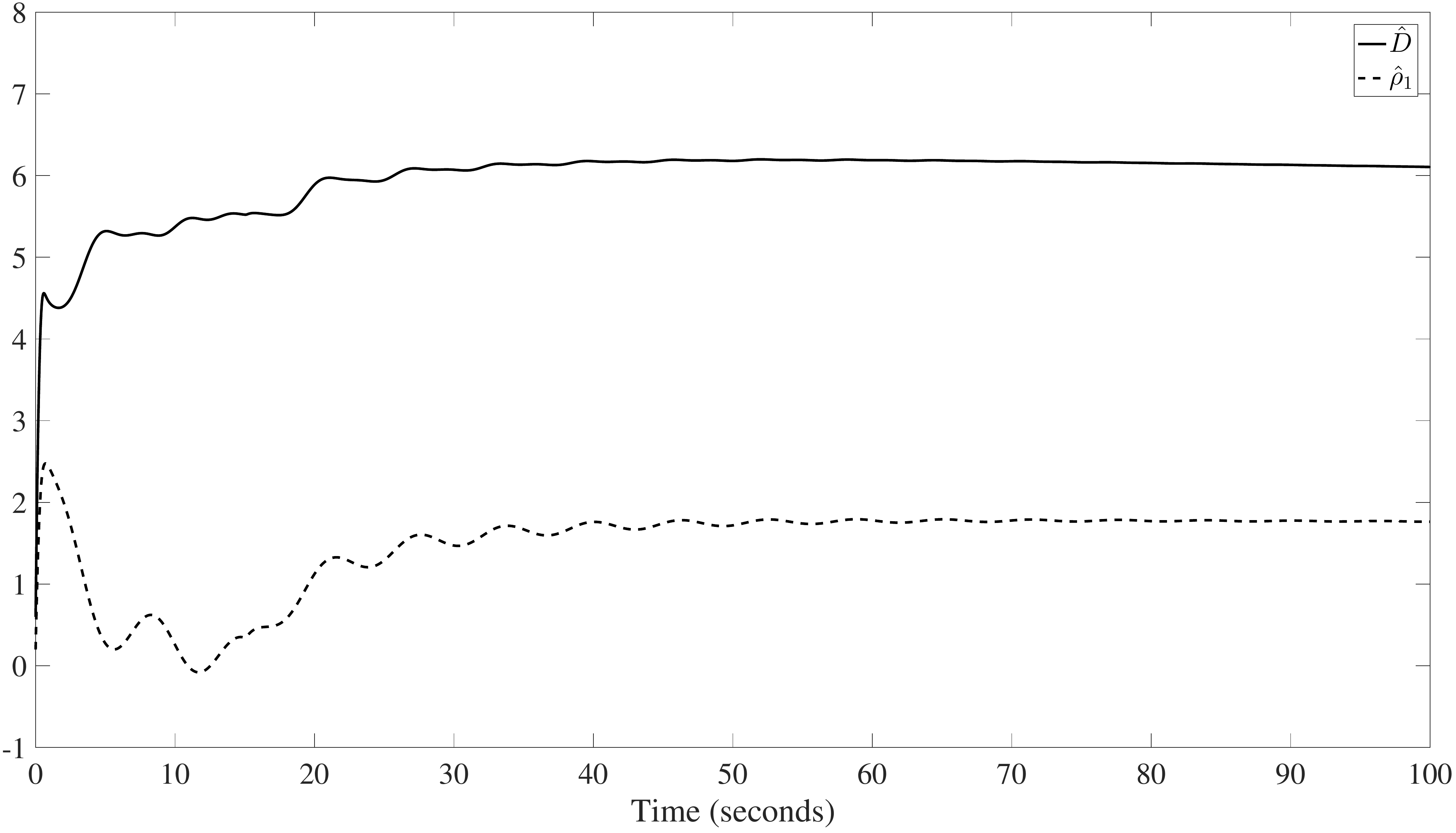}
\caption{The estimates $\hat{D}$ and $\hat{\rho}_1$.}
\label{fig:Dsrhos_d}
\end{minipage}
\end{tabular}
\end{figure}

\begin{figure}[h]
\begin{tabular}{cc}
\begin{minipage}[t]{0.48\linewidth}
\includegraphics[width=7.4cm,height=5.4cm]{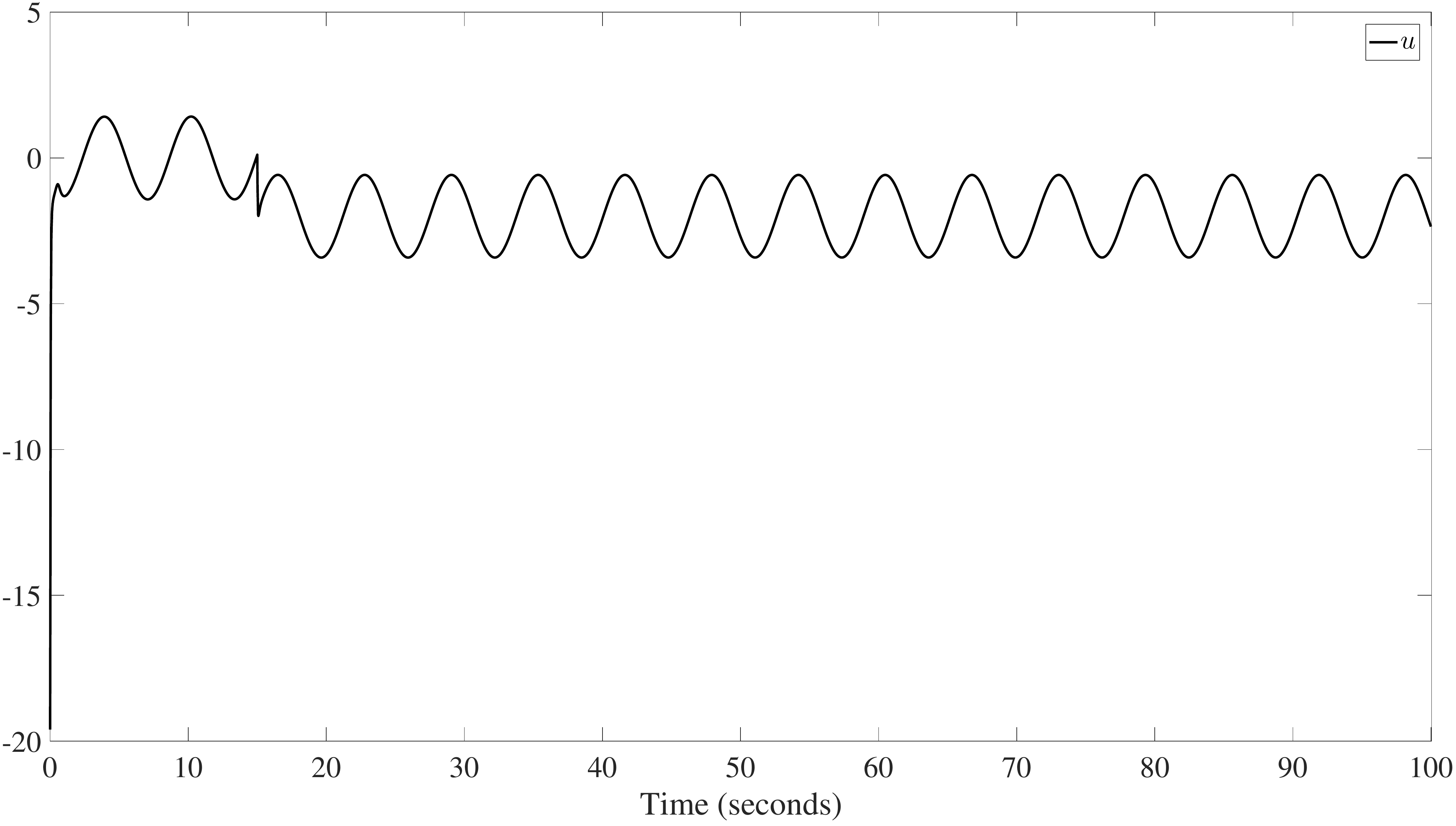}
\caption{The control input $u$.}
\label{fig:u_d}
\end{minipage}
\begin{minipage}[t]{0.48\linewidth}
\includegraphics[width=7.4cm,height=5.4cm]{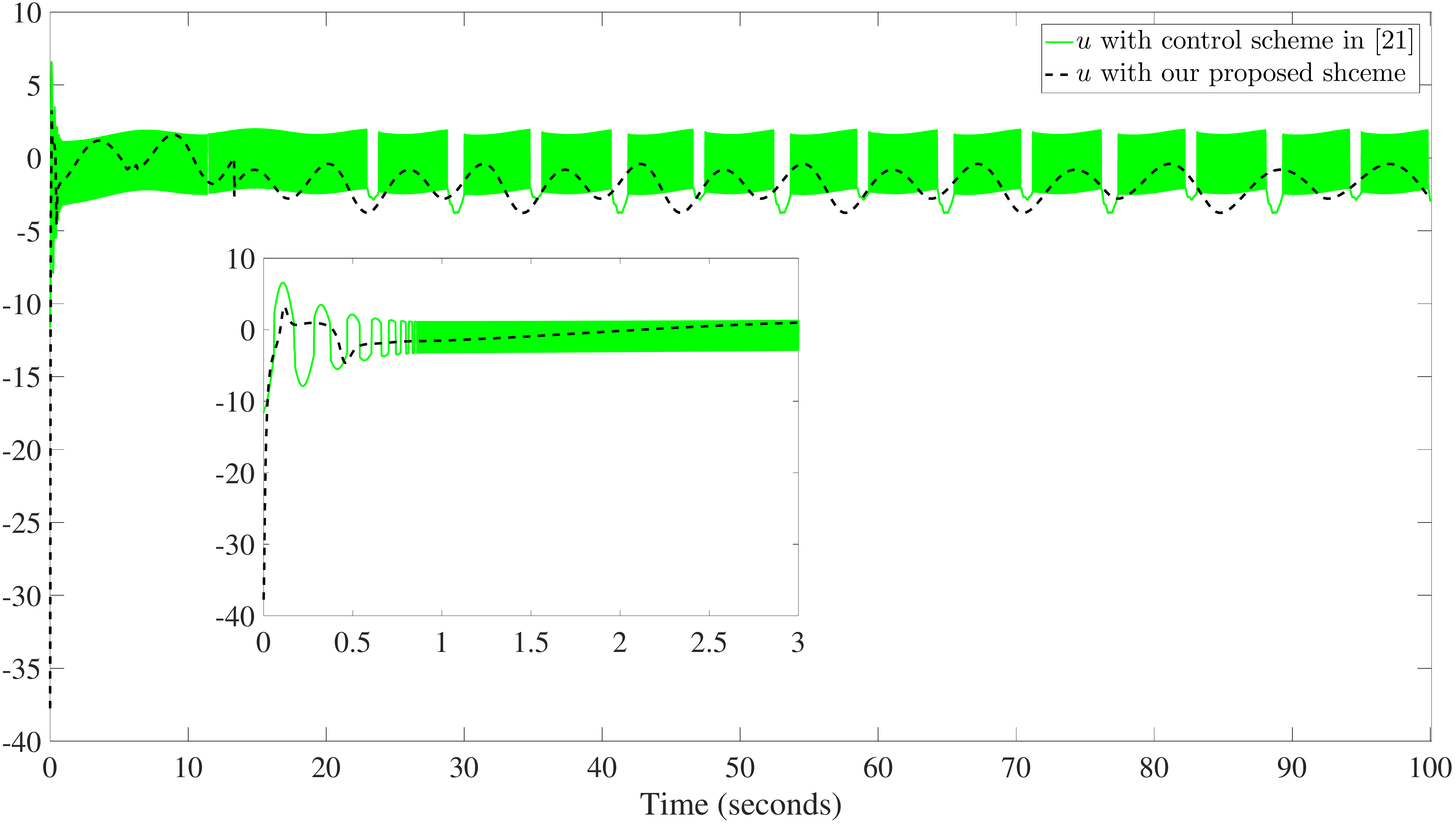}
\caption{The control input $u$ with control scheme involving $\mathrm{sign}(\cdot)$ in \cite{liu2017adaptive} and with scheme in this paper.}
\label{fig:comu}
\end{minipage}
\end{tabular}
\end{figure}

Firstly, the system is simulated when no control is applied, i.e. $u(t)=0$, and the result is given in Fig.~\ref{fig:x1x2nou}. As observed from the figure, in this case, the system (\ref{eq:eq2nd}) is unstable in the presence of disturbance $d(t)$. To stabilize the system, we apply our proposed adaptive control scheme presented in Section \ref{sec:controller}. The designed control parameters are selected as $c_1=30,\;c_2=1,\;\Gamma_1=\Gamma_2=2\mathit{I},\;\eta=4,\;\lambda_1=2$. The responses in this case are shown in Fig.~\ref{fig:yz2_d} to Fig.~\ref{fig:Dsrhos_d}. It could be found from Fig.~\ref{fig:yz2_d} that the system output $y$ and the virtual error $z_2$ are driven to $0$ eventually. Fig.~\ref{fig:thetas_d} and Fig.~\ref{fig:Dsrhos_d} show the estimations of uncertain parameters $\hat{\theta}_{v,1}$ and $\hat{\theta}_{v,2}$ and bound of external disturbance $\hat{D}$ as well as the bound of $\rho_1(t)$ accordingly. The control signal $u$ is shown in Fig.~\ref{fig:u_d}.

\begin{figure}[!htb]
\begin{tabular}{cc}
\begin{minipage}[t]{0.48\linewidth}
\includegraphics[width=7.4cm,height=5.4cm]{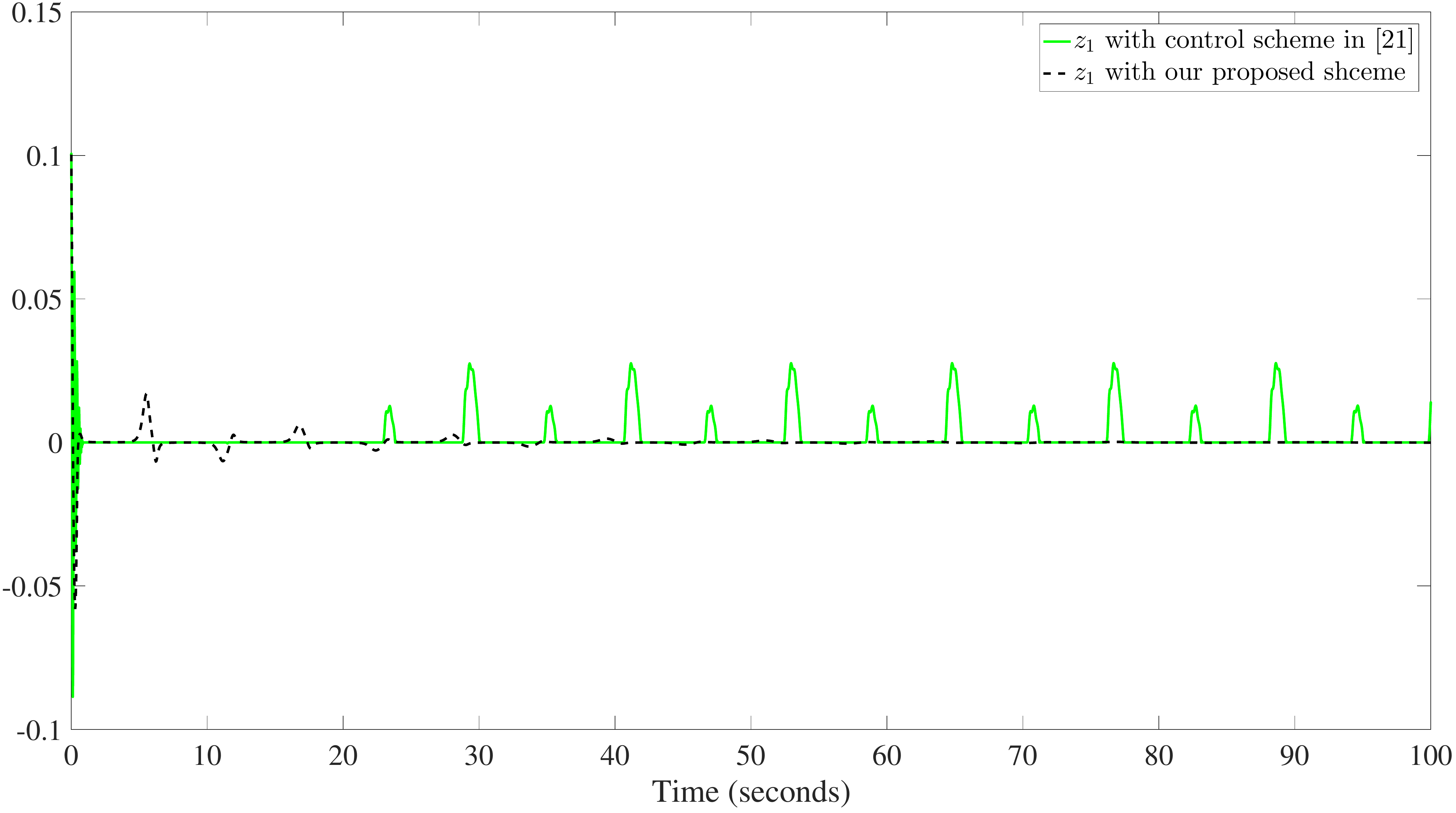}
\caption{The output tracking error $z_1$ with control scheme involving $\mathrm{sign}(\cdot)$ in \cite{liu2017adaptive} and with scheme in this paper.}
\label{fig:comz1}
\end{minipage}
\begin{minipage}[t]{0.48\linewidth}
\includegraphics[width=7.4cm,height=5.4cm]{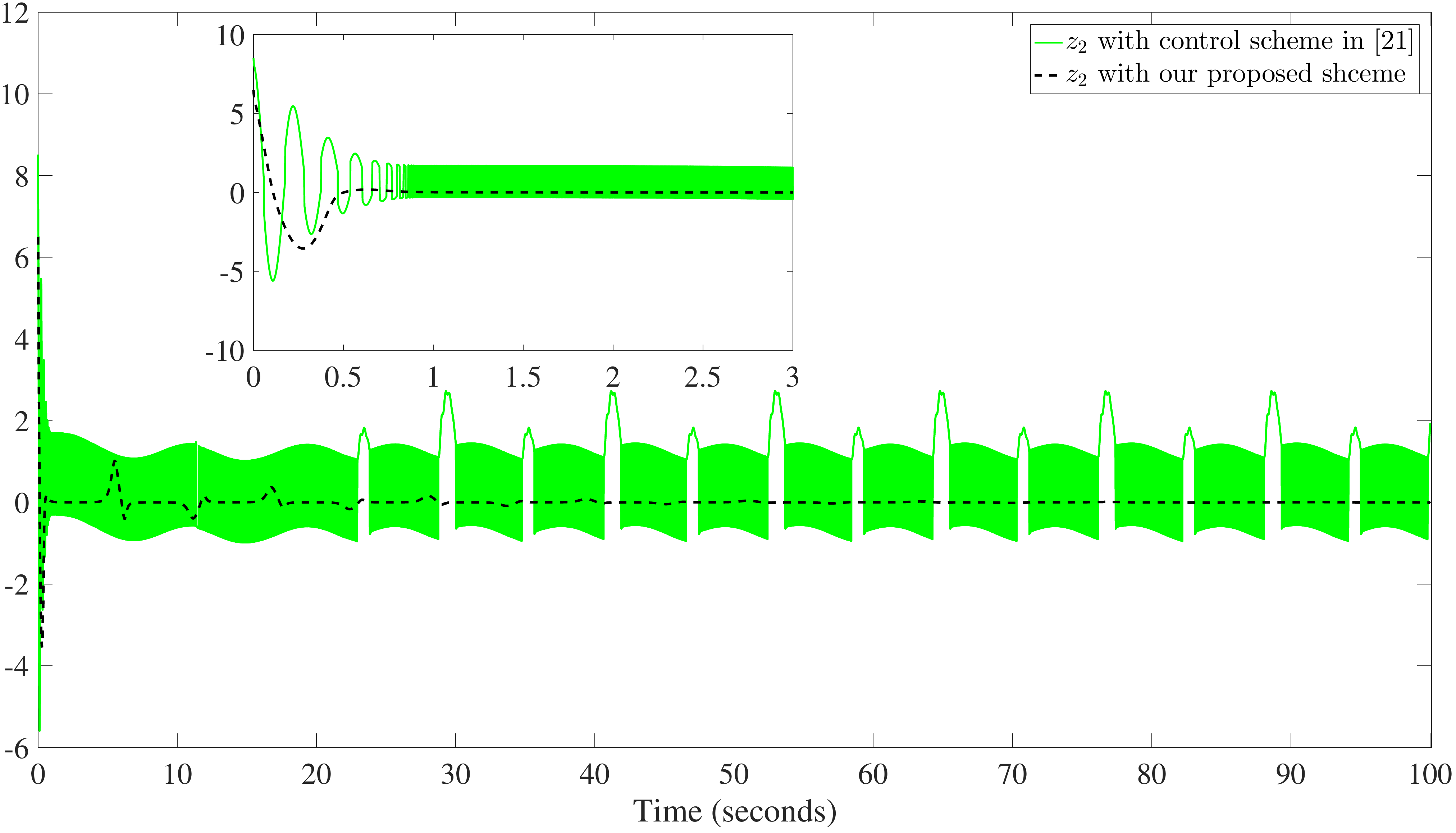}
\caption{The virtual error $z_2$ with control scheme involving $\mathrm{sign}(\cdot)$ in \cite{liu2017adaptive} and with scheme in this paper.}
\label{fig:comz2}
\end{minipage}
\end{tabular}
\end{figure}

\begin{figure}[!htb]
\begin{tabular}{cc}
\begin{minipage}[t]{0.48\linewidth}
\includegraphics[width=7.4cm,height=5.4cm]{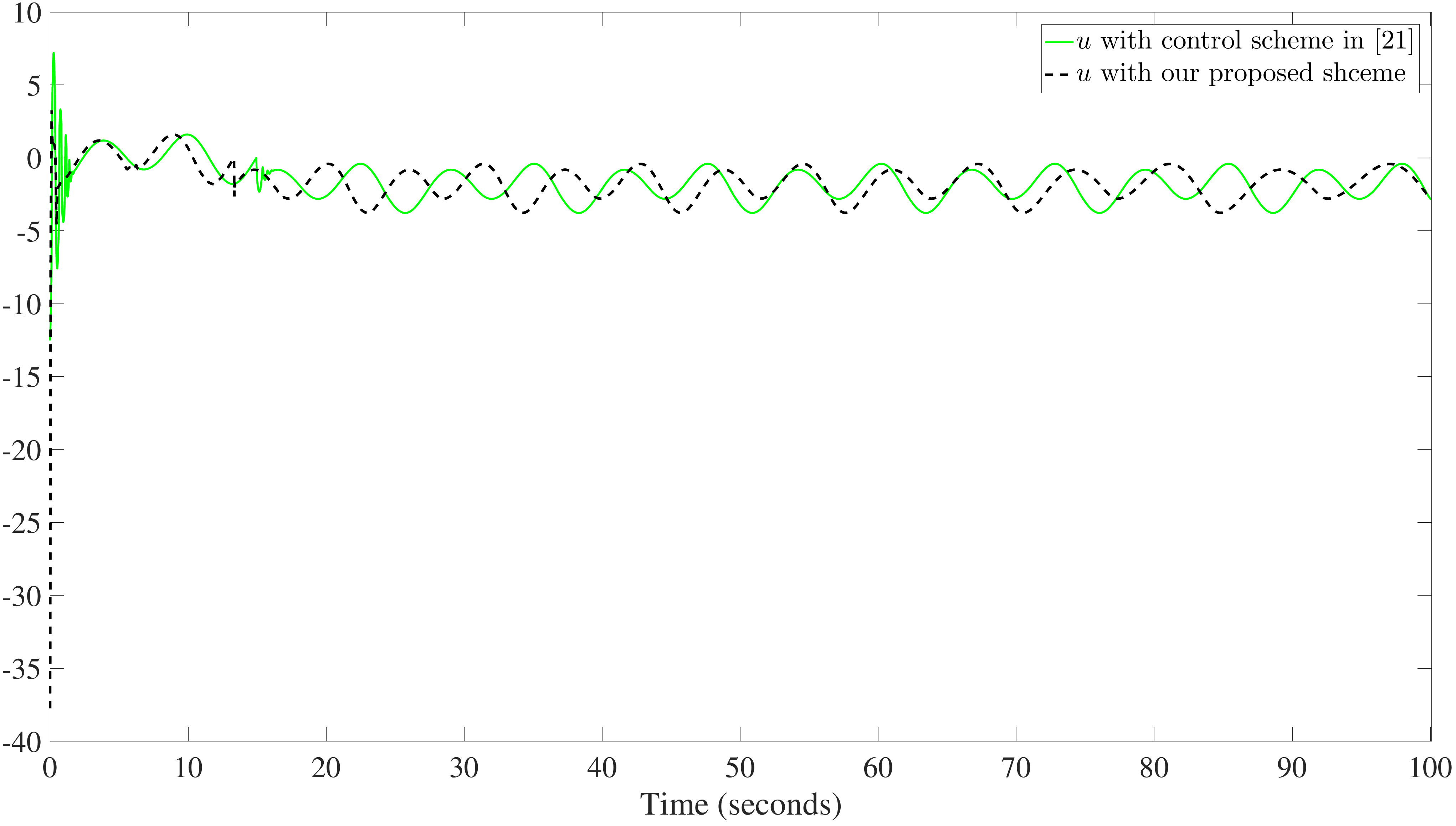}
\caption{The control input $u$ with control scheme involving $\mathrm{arctan}(10\cdot)$ in \cite{liu2017adaptive} and with scheme in this paper.}
\label{fig:ncomu}
\end{minipage}
\begin{minipage}[t]{0.48\linewidth}
\includegraphics[width=7.4cm,height=5.4cm]{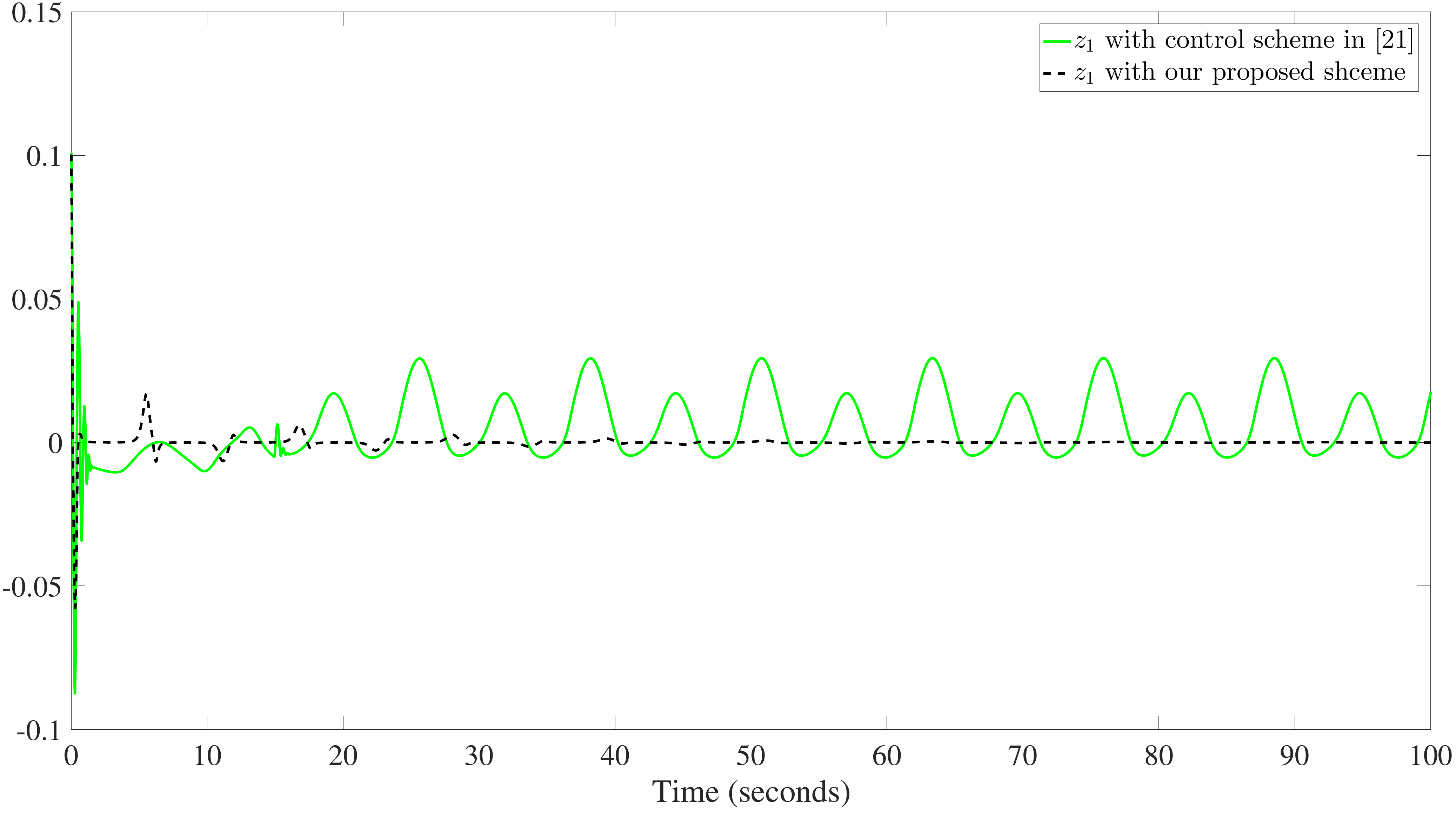}
\caption{The output tracking error $z_1$ with control scheme involving $\mathrm{arctan}(10\cdot)$ in \cite{liu2017adaptive} and with scheme in this paper.}
\label{fig:ncomz1}
\end{minipage}
\end{tabular}
\end{figure}

\begin{figure}[!htb]
\begin{tabular}{cc}
\begin{minipage}[t]{0.48\linewidth}
\includegraphics[width=7.4cm,height=5.4cm]{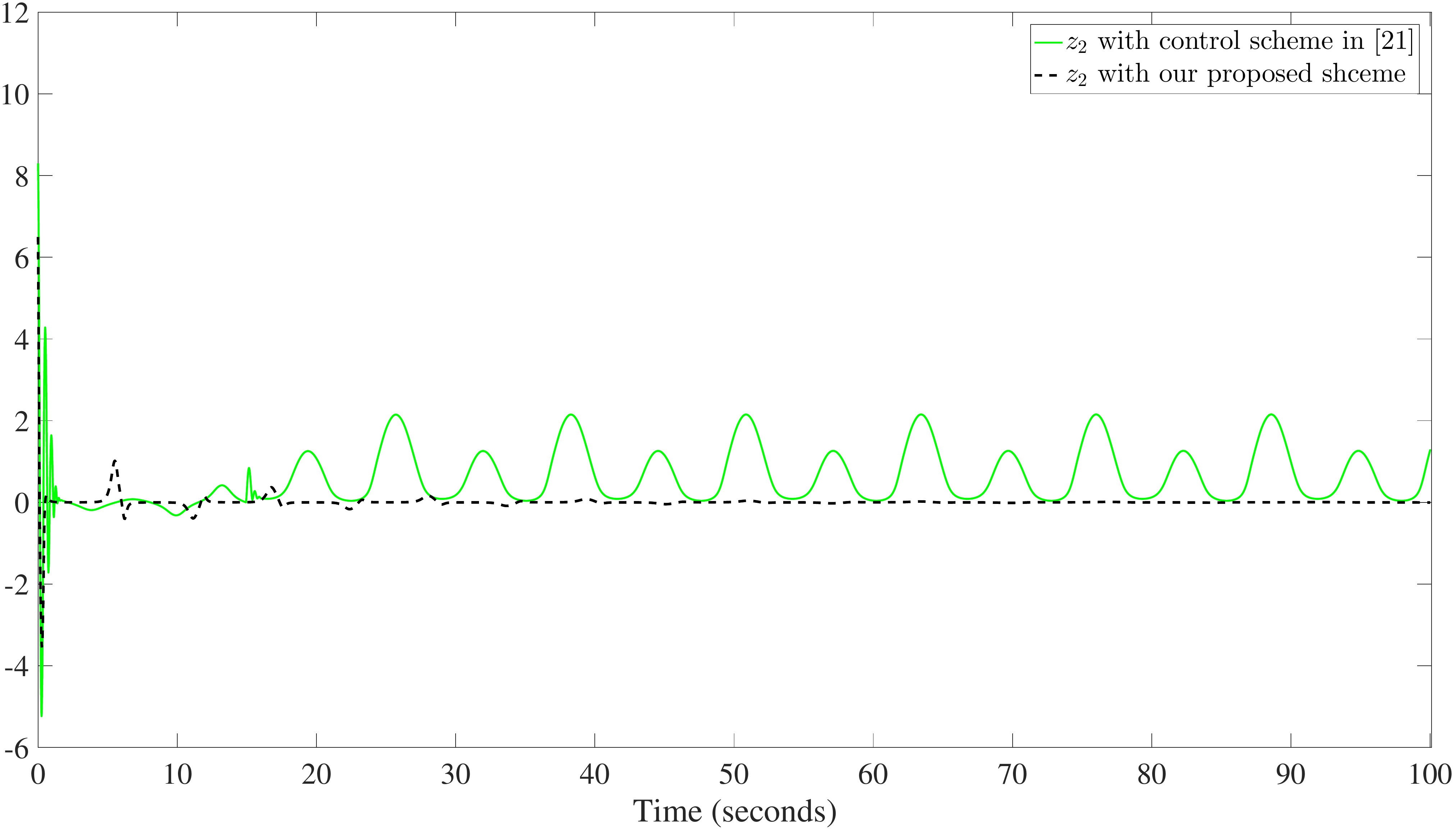}
\caption{The virtual error $z_2$ with control scheme involving $\mathrm{arctan}(10\cdot)$ in \cite{liu2017adaptive} and with scheme in this paper.}
\label{fig:ncomz2}
\end{minipage}
\begin{minipage}[t]{0.48\linewidth}
\includegraphics[width=7.8cm,height=5.4cm]{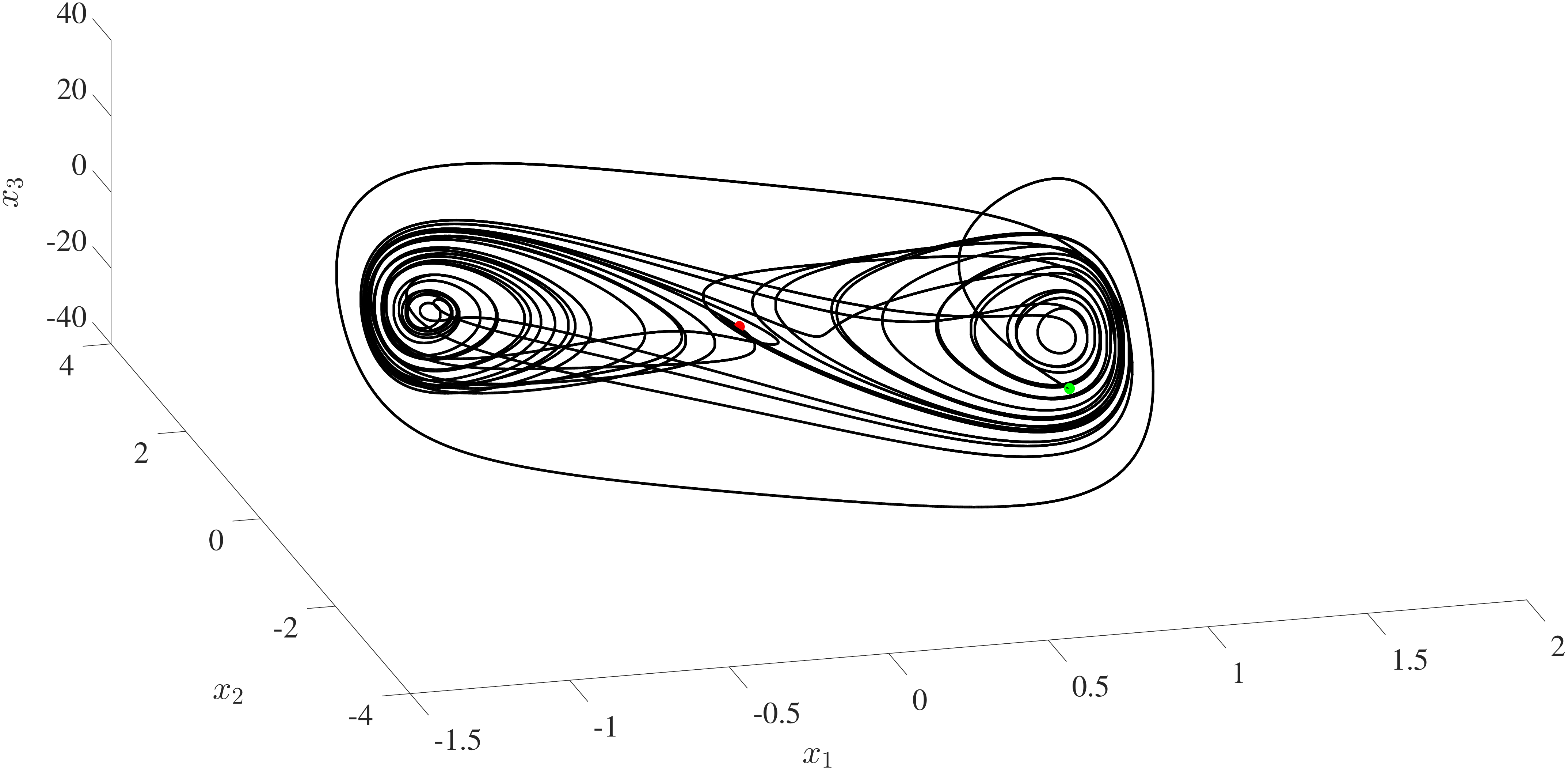}
\caption{The state variables of uncontrolled system (\ref{eq:eq3nd}) with \emph{green dot} and \emph{red dot} representing initial condition and the origin respectively.}
\label{fig:nx1x2x3nou100_d}
\end{minipage}
\end{tabular}
\end{figure}

To better illustrate the effectiveness of our proposed control algorithm, a comparative simulation study between the schemes in \cite{liu2017adaptive} and this paper is conducted under the same control objective that the output signal tracks a reference signal $r(t)=\mathrm{sin}(0.5t)$ while ensuring all the signals bounded. 
In \cite{liu2017adaptive}, fuzzy logic systems are employed to approximate unknown compounded nonlinear functions in the systems and also the fractional-order derivatives of the virtual control functions. 
The $\mathrm{sign}$ function is used to compensate for the effects caused by system uncertainties and approximated errors, thus causing chattering phenomenon. The comparison results are given in Fig.~\ref{fig:comu} to Fig.~\ref{fig:comz2}, from which we can observe that our proposed method can guarantee both the tracking error $z_1$ and virtual error $z_2$ converge to $0$ asymptotically without leading to chattering phenomenon in control signal $u$. On the contrary, not only the tracking error $z_1$ cannot go to $0$, but also chattering phenomena in the control signal $u$ and $z_2$ are caused by using control signal in \cite{liu2017adaptive}, which is also consistent with its theoretical results established.
Furthermore, it is proposed in \cite{liu2017adaptive} that the chattering phenomenon can be avoided by replacing $\mathrm{sign}(\cdot)$ with $\mathrm{arctan}(10\cdot)$ in the controller design, without theoretical analysis on stability and tracking performance provided. Comparison simulations are also carried out with such a replacement and the results can be found in Fig.~\ref{fig:ncomu} to Fig.~\ref{fig:ncomz2}. It can be seen from these figures that although the control signal in \cite{liu2017adaptive} becomes smooth, the tracking error $z_1$ and the virtual error $z_2$ can only be driven to small regions near $0$. However, by utilizing the control method proposed in this paper, both $z_1$ and $z_2$ can be asymptotically stabilized under smooth control signal $u$ with similar magnitude.

\vspace{1pt}

\vspace{1pt}

\subsection{An Example of Third-order System}

\begin{figure}[!htb]
\begin{tabular}{cc}
\begin{minipage}[t]{0.48\linewidth}
\includegraphics[width=7.8cm,height=5.4cm]{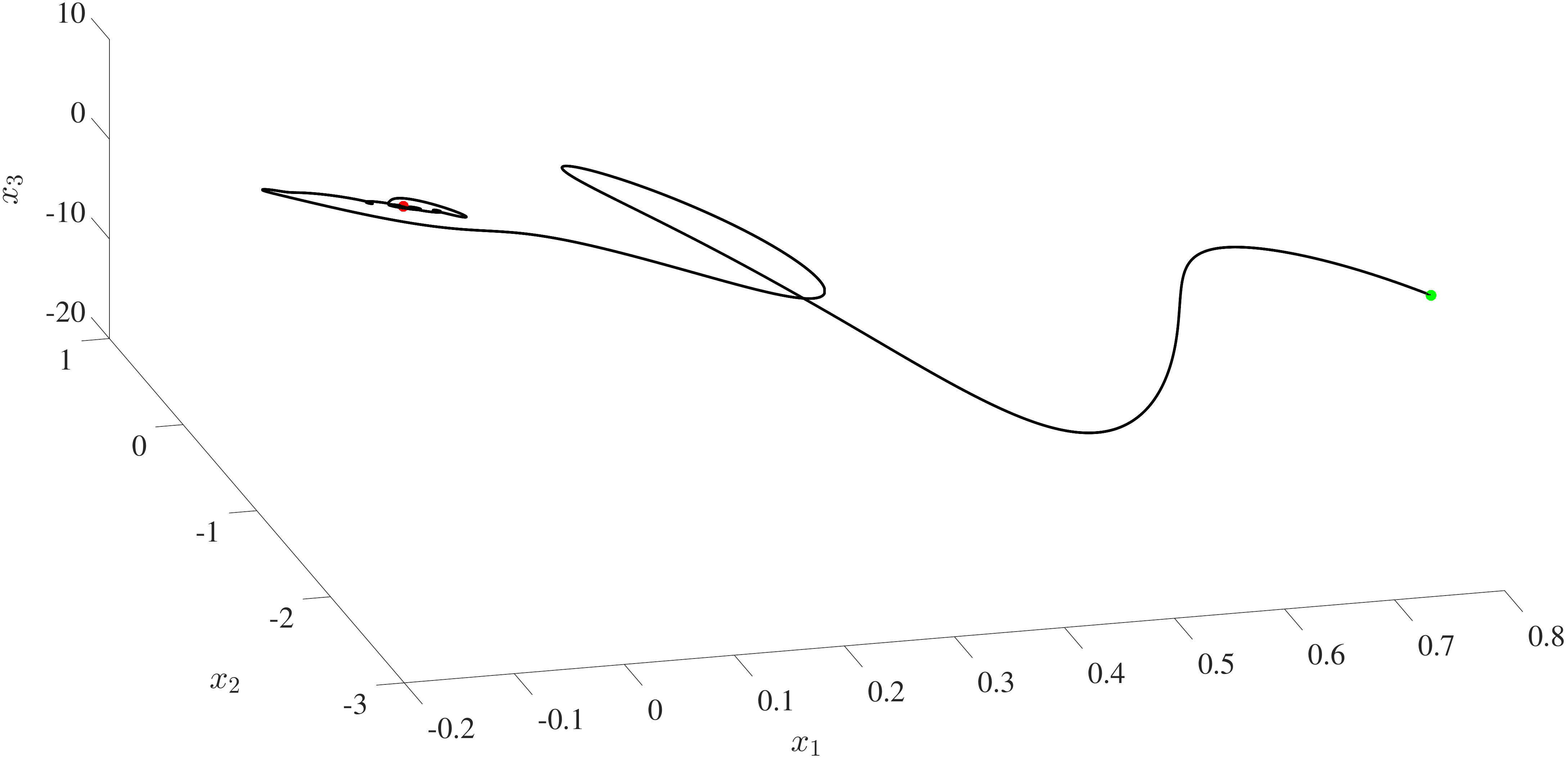}
\caption{The state variables of (\ref{eq:eq3nd}) under proposed control scheme with \emph{green dot} and \emph{red dot} representing initial condition and the origin respectively.}
\label{fig:nx1x2x3u_d}
\end{minipage}
\begin{minipage}[t]{0.48\linewidth}
\includegraphics[width=7.4cm,height=5.4cm]{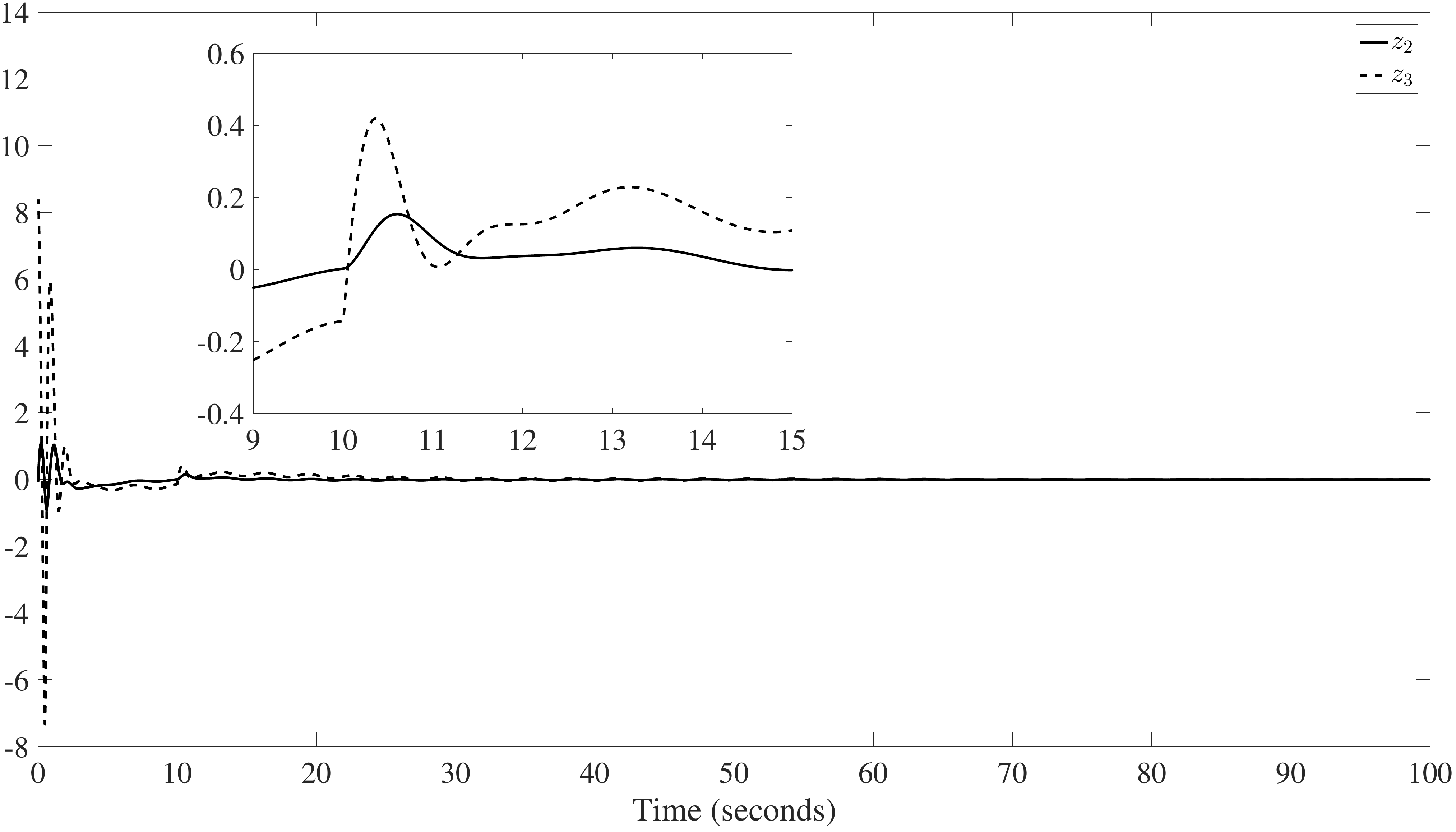}
\caption{The virtual errors $z_2$ and $z_3$.}
\label{fig:nz2z3_d}
\end{minipage}
\end{tabular}
\end{figure}

\begin{figure}[!htb]
\centering
\includegraphics[width=7.4cm,height=5.4cm]{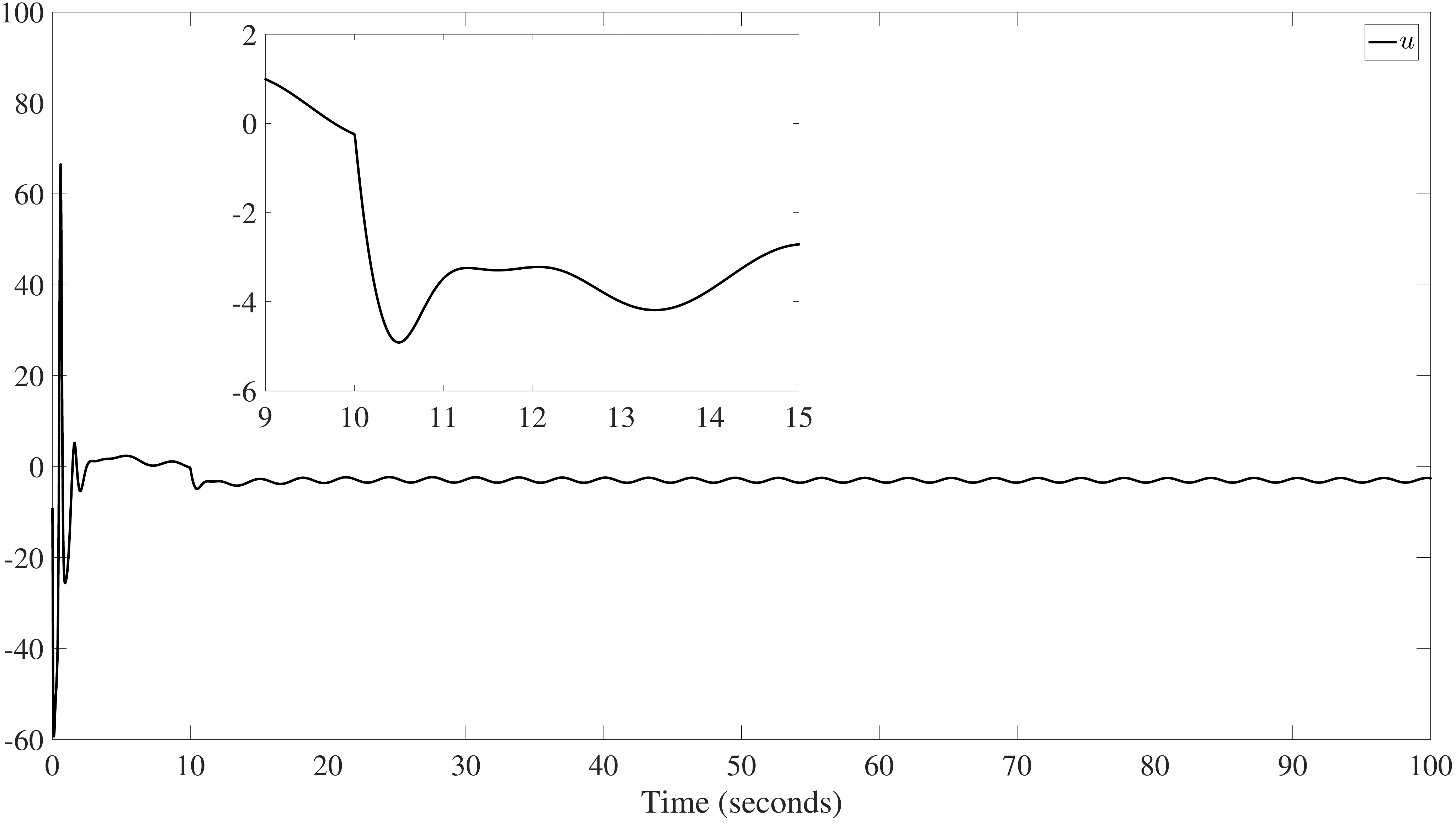}
\caption{The control input $u$ of (\ref{eq:eq3nd}).}
\label{fig:nu_d}
\end{figure}

Consider the following third-order fractional system which  is also known as Chua-Hartley's system \cite{petravs2008note}.

\begin{equation}
\begin{cases}
\begin{aligned}
{}_0^{C}\!{\mathcal{D}}_t^{\alpha}&x_1(t) = x_2(t)+\phi_1(x_1(t))\theta_1\\
{}_0^{C}\!{\mathcal{D}}_t^{\alpha}&x_2(t) = x_3(t)+\phi_2(x_1(t),x_2(t))\theta_2\\
{}_0^{C}\!{\mathcal{D}}_t^{\alpha}&x_{3}(t) = u(t)+\phi_{3}(x_1(t),x_2(t),x_3(t))\theta_{3}+d(t)\\
y(t)&=x_1(t)\\
\end{aligned}
\end{cases}
\label{eq:eq3nd}
\end{equation}
where $\alpha =0.98,\;\phi_1(x_1)=\frac{10}{7}(x_1-x_1^3),\;\phi_{2}(x_1,x_2)=10x_1-x_2,\;\phi_{3}(x_1,x_2,x_3)=-\frac{100}{7}x_2$. Suppose that $\theta_1=\theta_2=\theta_3=1$ and external time-varying disturbance $d(t)=0.5\mathrm{sin}(2t)+3\mathrm{U}(t-10)$, but they are unknown to controller design. The designed control parameters are selected as $c_1=c_2=c_3=2,\;\Gamma_1=\Gamma_2=\Gamma_3=0.1\mathit{I},\;\lambda_1=\lambda_2=1,\;\eta=0.1$. To make a comparison, we simulate the system without control and with our proposed control, under the same initial condition $x(0)= [0.8,\,-2,\,1]$. The behaviour of its state variables are shown in Fig.~\ref{fig:nx1x2x3nou100_d} and Fig.~\ref{fig:nx1x2x3u_d}, where the green dot indicates the initial state, respectively. As observed from Fig.~\ref{fig:nx1x2x3nou100_d}, the system without control exhibits chaotic phenomenon. From Fig.~\ref{fig:nx1x2x3u_d}, it is seen that our proposed control enables the chaotic behaviors of the original uncontrolled system to be removed. 
Moreover, from Fig.~\ref{fig:nx1x2x3u_d} and Fig.~\ref{fig:nz2z3_d}, the virtual errors as well as all the states will finally be driven to $0$ with control signal $u$ shown in Fig.~\ref{fig:nu_d}, which also reveals that the system is asymptotically stabilized. All these results illustrate the effectiveness of the proposed schemes and verify our theoretical results established.

\section{Conclusion}
\label{sec:conclusion}
In this paper, we propose a smooth adaptive backstepping control design scheme for a class of SISO commensurate fractional-order nonlinear systems in strict feedback form with uncertain system parameters and unknown external time-varying disturbance. It is proved that the resulting closed-loop system is globally asymptotically stable, even in the presence of arbitrary uncertainties and bounded disturbances. Simulation results also demonstrate the effectiveness in stabilizing unstable system and tracking reference signal with better performances compared to an existing scheme presented in \cite{liu2017adaptive}. For fractional order $\alpha > 1$, the control protocol in this paper is not theoretically shown effective, and therefore it is an interesting future research topic to extend our result to such systems.

\bibliographystyle{unsrt}  

\bibliography{references}

\end{document}